\documentclass[journal,10pt,twoside,twocolumn]{IEEEtran}

\usepackage{cite}
\usepackage[T1]{fontenc}
\usepackage{graphicx}

\usepackage{amssymb}
\usepackage{amsmath}
\usepackage{paralist}
\usepackage{subfigure}
\usepackage{booktabs} 
\usepackage{algorithm}
\usepackage{algorithmic}

\usepackage{amssymb}
\usepackage{amsfonts}
\usepackage{mathrsfs}
\usepackage{xspace}
\usepackage{bm}
\usepackage{upgreek}

\newcommand{\safemath}[2]{\newcommand{#1}{\ensuremath{#2}\xspace}}

\safemath{\bma}{\mathbf{a}}
\safemath{\bmb}{\mathbf{b}}
\safemath{\bmc}{\mathbf{c}}
\safemath{\bmd}{\mathbf{d}}
\safemath{\bme}{\mathbf{e}}
\safemath{\bmf}{\mathbf{f}}
\safemath{\bmg}{\mathbf{g}}
\safemath{\bmh}{\mathbf{h}}
\safemath{\bmi}{\mathbf{i}}
\safemath{\bmj}{\mathbf{j}}
\safemath{\bmk}{\mathbf{k}}
\safemath{\bml}{\mathbf{l}}
\safemath{\bmm}{\mathbf{m}}
\safemath{\bmn}{\mathbf{n}}
\safemath{\bmo}{\mathbf{o}}
\safemath{\bmp}{\mathbf{p}}
\safemath{\bmq}{\mathbf{q}}
\safemath{\bmr}{\mathbf{r}}
\safemath{\bms}{\mathbf{s}}
\safemath{\bmt}{\mathbf{t}}
\safemath{\bmu}{\mathbf{u}}
\safemath{\bmv}{\mathbf{v}}
\safemath{\bmw}{\mathbf{w}}
\safemath{\bmx}{\mathbf{x}}
\safemath{\bmy}{\mathbf{y}}
\safemath{\bmz}{\mathbf{z}}
\safemath{\bmzero}{\mathbf{0}}
\safemath{\bmone}{\mathbf{1}}

\bmdefine{\biad}{a}
\bmdefine{\bibd}{b}
\bmdefine{\bicd}{c}
\bmdefine{\bidd}{d}
\bmdefine{\bied}{e}
\bmdefine{\bifd}{f}
\bmdefine{\bigd}{g}
\bmdefine{\bihd}{h}
\bmdefine{\biid}{i}
\bmdefine{\bijd}{j}
\bmdefine{\bikd}{k}
\bmdefine{\bild}{l}
\bmdefine{\bimd}{m}
\bmdefine{\bind}{n}
\bmdefine{\biod}{o}
\bmdefine{\bipd}{p}
\bmdefine{\biqd}{q}
\bmdefine{\bird}{r}
\bmdefine{\bisd}{s}
\bmdefine{\bitd}{t}
\bmdefine{\biud}{u}
\bmdefine{\bivd}{v}
\bmdefine{\biwd}{w}
\bmdefine{\bixd}{x}
\bmdefine{\biyd}{y}
\bmdefine{\bizd}{z}

\bmdefine{\bixid}{\xi}
\bmdefine{\bilambdad}{\lambda}
\bmdefine{\bimud}{\mu}
\bmdefine{\bithetad}{\theta}
\bmdefine{\biphid}{\phi}
\bmdefine{\bideltad}{\delta}

\safemath{\bmia}{\biad}
\safemath{\bmib}{\bibd}
\safemath{\bmic}{\bicd}
\safemath{\bmid}{\bidd}
\safemath{\bmie}{\bied}
\safemath{\bmif}{\bifd}
\safemath{\bmig}{\bigd}
\safemath{\bmih}{\bihd}
\safemath{\bmii}{\biid}
\safemath{\bmij}{\bijd}
\safemath{\bmik}{\bikd}
\safemath{\bmil}{\bild}
\safemath{\bmim}{\bimd}
\safemath{\bmin}{\bind}
\safemath{\bmio}{\biod}
\safemath{\bmip}{\bipd}
\safemath{\bmiq}{\biqd}
\safemath{\bmir}{\bird}
\safemath{\bmis}{\bisd}
\safemath{\bmit}{\bitd}
\safemath{\bmiu}{\biud}
\safemath{\bmiv}{\bivd}
\safemath{\bmiw}{\biwd}
\safemath{\bmix}{\bixd}
\safemath{\bmiy}{\biyd}
\safemath{\bmiz}{\bizd}

\safemath{\bmxi}{\bixid}
\safemath{\bmlambda}{\bilambdad}
\safemath{\bmmu}{\bimud}
\safemath{\bmtheta}{\bithetad}
\safemath{\bmphi}{\biphid}
\safemath{\bmdelta}{\bideltad}

\safemath{\bA}{\mathbf{A}}
\safemath{\bB}{\mathbf{B}}
\safemath{\bC}{\mathbf{C}}
\safemath{\bD}{\mathbf{D}}
\safemath{\bE}{\mathbf{E}}
\safemath{\bF}{\mathbf{F}}
\safemath{\bG}{\mathbf{G}}
\safemath{\bH}{\mathbf{H}}
\safemath{\bI}{\mathbf{I}}
\safemath{\bJ}{\mathbf{J}}
\safemath{\bK}{\mathbf{K}}
\safemath{\bL}{\mathbf{L}}
\safemath{\bM}{\mathbf{M}}
\safemath{\bN}{\mathbf{N}}
\safemath{\bO}{\mathbf{O}}
\safemath{\bP}{\mathbf{P}}
\safemath{\bQ}{\mathbf{Q}}
\safemath{\bR}{\mathbf{R}}
\safemath{\bS}{\mathbf{S}}
\safemath{\bT}{\mathbf{T}}
\safemath{\bU}{\mathbf{U}}
\safemath{\bV}{\mathbf{V}}
\safemath{\bW}{\mathbf{W}}
\safemath{\bX}{\mathbf{X}}
\safemath{\bY}{\mathbf{Y}}
\safemath{\bZ}{\mathbf{Z}}

\safemath{\bZero}{\mathbf{0}}
\safemath{\bOne}{\mathbf{1}}
\safemath{\bDelta}{\mathbf{\Delta}}
\safemath{\bLambda}{\mathbf{\UpLambda}}
\safemath{\bPhi}{\mathbf{\Upphi}}
\safemath{\bSigma}{\mathbf{\Upsigma}}
\safemath{\bOmega}{\mathbf{\Upomega}}
\safemath{\bTheta}{\mathbf{\Uptheta}}

\bmdefine{\biAd}{A}
\bmdefine{\biBd}{B}
\bmdefine{\biCd}{C}
\bmdefine{\biDd}{D}
\bmdefine{\biEd}{E}
\bmdefine{\biFd}{F}
\bmdefine{\biGd}{G}
\bmdefine{\biHd}{H}
\bmdefine{\biId}{I}
\bmdefine{\biJd}{J}
\bmdefine{\biKd}{K}
\bmdefine{\biLd}{L}
\bmdefine{\biMd}{M}
\bmdefine{\biOd}{N}
\bmdefine{\biPd}{O}
\bmdefine{\biQd}{P}
\bmdefine{\biRd}{R}
\bmdefine{\biSd}{S}
\bmdefine{\biTd}{T}
\bmdefine{\biUd}{U}
\bmdefine{\biVd}{V}
\bmdefine{\biWd}{W}
\bmdefine{\biXd}{X}
\bmdefine{\biYd}{Y}
\bmdefine{\biZd}{Z}

\bmdefine{\biDelta}{\Delta}
\bmdefine{\biLambda}{\Lambda}
\bmdefine{\biPhi}{\Phi}
\bmdefine{\biSigma}{\Sigma}
\bmdefine{\biOmega}{\Omega}
\bmdefine{\biTheta}{\Theta}

\safemath{\bimA}{\biAd}
\safemath{\bimB}{\biBd}
\safemath{\bimC}{\biCd}
\safemath{\bimD}{\biDd}
\safemath{\bimE}{\biEd}
\safemath{\bimF}{\biFd}
\safemath{\bimG}{\biGd}
\safemath{\bimH}{\biHd}
\safemath{\bimI}{\biId}
\safemath{\bimJ}{\biJd}
\safemath{\bimK}{\biKd}
\safemath{\bimL}{\biLd}
\safemath{\bimM}{\biMd}
\safemath{\bimN}{\biNd}
\safemath{\bimO}{\biOd}
\safemath{\bimP}{\biPd}
\safemath{\bimQ}{\biQd}
\safemath{\bimR}{\biRd}
\safemath{\bimS}{\biSd}
\safemath{\bimT}{\biTd}
\safemath{\bimU}{\biUd}
\safemath{\bimV}{\biVd}
\safemath{\bimW}{\biWd}
\safemath{\bimX}{\biXd}
\safemath{\bimY}{\biYd}
\safemath{\bimZ}{\biZd}

\safemath{\bimDelta}{\biDelta}
\safemath{\bimLambda}{\biLambda}
\safemath{\bimPhi}{\biPhi}
\safemath{\bimSigma}{\biSigma}
\safemath{\bimOmega}{\biOmega}
\safemath{\bimTheta}{\biTheta}

\safemath{\setA}{\mathcal{A}}
\safemath{\setB}{\mathcal{B}}
\safemath{\setC}{\mathcal{C}}
\safemath{\setD}{\mathcal{D}}
\safemath{\setE}{\mathcal{E}}
\safemath{\setF}{\mathcal{F}}
\safemath{\setG}{\mathcal{G}}
\safemath{\setH}{\mathcal{H}}
\safemath{\setI}{\mathcal{I}}
\safemath{\setJ}{\mathcal{J}}
\safemath{\setK}{\mathcal{K}}
\safemath{\setL}{\mathcal{L}}
\safemath{\setM}{\mathcal{M}}
\safemath{\setN}{\mathcal{N}}
\safemath{\setO}{\mathcal{O}}
\safemath{\setP}{\mathcal{P}}
\safemath{\setQ}{\mathcal{Q}}
\safemath{\setR}{\mathcal{R}}
\safemath{\setS}{\mathcal{S}}
\safemath{\setT}{\mathcal{T}}
\safemath{\setU}{\mathcal{U}}
\safemath{\setV}{\mathcal{V}}
\safemath{\setW}{\mathcal{W}}
\safemath{\setX}{\mathcal{X}}
\safemath{\setY}{\mathcal{Y}}
\safemath{\setZ}{\mathcal{Z}}
\safemath{\emptySet}{\varnothing}

\safemath{\colA}{\mathscr{A}}
\safemath{\colB}{\mathscr{B}}
\safemath{\colC}{\mathscr{C}}
\safemath{\colD}{\mathscr{D}}
\safemath{\colE}{\mathscr{E}}
\safemath{\colF}{\mathscr{F}}
\safemath{\colG}{\mathscr{G}}
\safemath{\colH}{\mathscr{H}}
\safemath{\colI}{\mathscr{I}}
\safemath{\colJ}{\mathscr{J}}
\safemath{\colK}{\mathscr{K}}
\safemath{\colL}{\mathscr{L}}
\safemath{\colM}{\mathscr{M}}
\safemath{\colN}{\mathscr{N}}
\safemath{\colO}{\mathscr{O}}
\safemath{\colP}{\mathscr{P}}
\safemath{\colQ}{\mathscr{Q}}
\safemath{\colR}{\mathscr{R}}
\safemath{\colS}{\mathscr{S}}
\safemath{\colT}{\mathscr{T}}
\safemath{\colU}{\mathscr{U}}
\safemath{\colV}{\mathscr{V}}
\safemath{\colW}{\mathscr{W}}
\safemath{\colX}{\mathscr{X}}
\safemath{\colY}{\mathscr{Y}}
\safemath{\colZ}{\mathscr{Z}}

\safemath{\opA}{\mathbb{A}}
\safemath{\opB}{\mathbb{B}}
\safemath{\opC}{\mathbb{C}}
\safemath{\opD}{\mathbb{D}}
\safemath{\opE}{\mathbb{E}}
\safemath{\opF}{\mathbb{F}}
\safemath{\opG}{\mathbb{G}}
\safemath{\opH}{\mathbb{H}}
\safemath{\opI}{\mathbb{I}}
\safemath{\opJ}{\mathbb{J}}
\safemath{\opK}{\mathbb{K}}
\safemath{\opL}{\mathbb{L}}
\safemath{\opM}{\mathbb{M}}
\safemath{\opN}{\mathbb{N}}
\safemath{\opO}{\mathbb{O}}
\safemath{\opP}{\mathbb{P}}
\safemath{\opQ}{\mathbb{Q}}
\safemath{\opR}{\mathbb{R}}
\safemath{\opS}{\mathbb{S}}
\safemath{\opT}{\mathbb{T}}
\safemath{\opU}{\mathbb{U}}
\safemath{\opV}{\mathbb{V}}
\safemath{\opW}{\mathbb{W}}
\safemath{\opX}{\mathbb{X}}
\safemath{\opY}{\mathbb{Y}}
\safemath{\opZ}{\mathbb{Z}}
\safemath{\opZero}{\mathbb{O}}
\safemath{\identityop}{\opI}

\safemath{\veca}{\bma}
\safemath{\vecb}{\bmb}
\safemath{\vecc}{\bmc}
\safemath{\vecd}{\bmd}
\safemath{\vece}{\bme}
\safemath{\vecf}{\bmf}
\safemath{\vecg}{\bmg}
\safemath{\vech}{\bmh}
\safemath{\veci}{\bmi}
\safemath{\vecj}{\bmj}
\safemath{\veck}{\bmk}
\safemath{\vecl}{\bml}
\safemath{\vecm}{\bmm}
\safemath{\vecn}{\bmn}
\safemath{\veco}{\bmo}
\safemath{\vecp}{\bmp}
\safemath{\vecq}{\bmq}
\safemath{\vecr}{\bmr}
\safemath{\vecs}{\bms}
\safemath{\vect}{\bmt}
\safemath{\vecu}{\bmu}
\safemath{\vecv}{\bmv}
\safemath{\vecw}{\bmw}
\safemath{\vecx}{\bmx}
\safemath{\vecy}{\bmy}
\safemath{\vecz}{\bmz}

\safemath{\veczero}{\bmzero}
\safemath{\vecone}{\bmone}
\safemath{\vecxi}{\bmxi}
\safemath{\veclambda}{\bmlambda}
\safemath{\vecmu}{\bmmu}
\safemath{\vectheta}{\bmtheta}
\safemath{\vecphi}{\bmphi}
\safemath{\vecdelta}{\bmdelta}

\safemath{\matA}{\bA}
\safemath{\matB}{\bB}
\safemath{\matC}{\bC}
\safemath{\matD}{\bD}
\safemath{\matE}{\bE}
\safemath{\matF}{\bF}
\safemath{\matG}{\bG}
\safemath{\matH}{\bH}
\safemath{\matI}{\bI}
\safemath{\matJ}{\bJ}
\safemath{\matK}{\bK}
\safemath{\matL}{\bL}
\safemath{\matM}{\bM}
\safemath{\matN}{\bN}
\safemath{\matO}{\bO}
\safemath{\matP}{\bP}
\safemath{\matQ}{\bQ}
\safemath{\matR}{\bR}
\safemath{\matS}{\bS}
\safemath{\matT}{\bT}
\safemath{\matU}{\bU}
\safemath{\matV}{\bV}
\safemath{\matW}{\bW}
\safemath{\matX}{\bX}
\safemath{\matY}{\bY}
\safemath{\matZ}{\bZ}
\safemath{\matzero}{\bmzero}

\safemath{\matDelta}{\bDelta}
\safemath{\matLambda}{\bLambda}
\safemath{\matPhi}{\bPhi}
\safemath{\matSigma}{\bSigma}
\safemath{\matOmega}{\bOmega}
\safemath{\matTheta}{\bTheta}

\safemath{\matidentity}{\matI}
\safemath{\matone}{\matO}

\safemath{\rnda}{A}
\safemath{\rndb}{B}
\safemath{\rndc}{C}
\safemath{\rndd}{D}
\safemath{\rnde}{E}
\safemath{\rndf}{F}
\safemath{\rndg}{G}
\safemath{\rndh}{H}
\safemath{\rndi}{I}
\safemath{\rndj}{J}
\safemath{\rndk}{K}
\safemath{\rndl}{L}
\safemath{\rndm}{M}
\safemath{\rndn}{N}
\safemath{\rndo}{O}
\safemath{\rndp}{P}
\safemath{\rndq}{Q}
\safemath{\rndr}{R}
\safemath{\rnds}{S}
\safemath{\rndt}{T}
\safemath{\rndu}{U}
\safemath{\rndv}{V}
\safemath{\rndw}{W}
\safemath{\rndx}{X}
\safemath{\rndy}{Y}
\safemath{\rndz}{Z}

\safemath{\rveca}{\bimA}
\safemath{\rvecb}{\bimB}
\safemath{\rvecc}{\bimC}
\safemath{\rvecd}{\bimD}
\safemath{\rvece}{\bimE}
\safemath{\rvecf}{\bimF}
\safemath{\rvecg}{\bimG}
\safemath{\rvech}{\bimH}
\safemath{\rveci}{\bimI}
\safemath{\rvecj}{\bimJ}
\safemath{\rveck}{\bimK}
\safemath{\rvecl}{\bimL}
\safemath{\rvecm}{\bimM}
\safemath{\rvecn}{\bimN}
\safemath{\rveco}{\bomO}
\safemath{\rvecp}{\bimP}
\safemath{\rvecq}{\bimQ}
\safemath{\rvecr}{\bimR}
\safemath{\rvecs}{\bimS}
\safemath{\rvect}{\bimT}
\safemath{\rvecu}{\bimU}
\safemath{\rvecv}{\bimV}
\safemath{\rvecw}{\bimW}
\safemath{\rvecx}{\bimX}
\safemath{\rvecy}{\bimY}
\safemath{\rvecz}{\bimZ}

\safemath{\rvecxi}{\bmxi}
\safemath{\rveclambda}{\bmlambda}
\safemath{\rvecmu}{\bmmu}
\safemath{\rvectheta}{\bmtheta}
\safemath{\rvecphi}{\bmphi}

\safemath{\rmatA}{\bimA}
\safemath{\rmatB}{\bimB}
\safemath{\rmatC}{\bimC}
\safemath{\rmatD}{\bimD}
\safemath{\rmatE}{\bimE}
\safemath{\rmatF}{\bimF}
\safemath{\rmatG}{\bimG}
\safemath{\rmatH}{\bimH}
\safemath{\rmatI}{\bimI}
\safemath{\rmatJ}{\bimJ}
\safemath{\rmatK}{\bimK}
\safemath{\rmatL}{\bimL}
\safemath{\rmatM}{\bimM}
\safemath{\rmatN}{\bimN}
\safemath{\rmatO}{\bimO}
\safemath{\rmatP}{\bimP}
\safemath{\rmatQ}{\bimQ}
\safemath{\rmatR}{\bimR}
\safemath{\rmatS}{\bimS}
\safemath{\rmatT}{\bimT}
\safemath{\rmatU}{\bimU}
\safemath{\rmatV}{\bimV}
\safemath{\rmatW}{\bimW}
\safemath{\rmatX}{\bimX}
\safemath{\rmatY}{\bimY}
\safemath{\rmatZ}{\bimZ}

\safemath{\rmatDelta}{\bimDelta}
\safemath{\rmatLambda}{\bimLambda}
\safemath{\rmatPhi}{\bimPhi}
\safemath{\rmatSigma}{\bimSigma}
\safemath{\rmatOmega}{\bimOmega}
\safemath{\rmatTheta}{\bimTheta}

\usepackage{amssymb}
\usepackage{amsfonts}
\usepackage{mathrsfs}
\usepackage{xspace}
\usepackage{bm}
\usepackage{fancyref}
\usepackage{textcomp}

\usepackage{multirow}
\usepackage{stmaryrd}

\newenvironment{textbmatrix}{	\setlength{\arraycolsep}{2.5pt}%
								\big[\begin{matrix}}{\end{matrix}\big]%
								\raisebox{0.08ex}{\vphantom{M}}}

\def\be{\begin{equation}}
\def\ee{\end{equation}}
\def\een{\nonumber \end{equation}}
\def\mat{\begin{bmatrix}}
\def\emat{\end{bmatrix}}
\def\btm{\begin{textbmatrix}}
\def\etm{\end{textbmatrix}}

\def\ba#1\ea{\begin{align}#1\end{align}}
\def\bas#1\eas{\begin{align*}#1\end{align*}}
\def\bs#1\es{\begin{split}#1\end{split}} 
\def\bg#1\eg{\begin{gather}#1\end{gather}}
\def\bml#1\eml{\begin{multline}#1\end{multline}}
\def\bi#1\ei{\begin{itemize}#1\end{itemize}}

\newcommand{\lefto}{\mathopen{}\left}

\DeclareMathOperator*{\argmin}{arg\;min}		
\DeclareMathOperator{\Prob}{\opP}			
\DeclareMathOperator{\Exop}{\opE}			
\DeclareMathOperator{\grad}{\nabla}			

\newcommand{\abs}[1]{\lefto\lvert#1\right\rvert}		

\newcommand{\vecnorm}[1]{\lefto\lVert#1\right\rVert}		
\newcommand{\herm}[1]{\ensuremath{#1^{H}}} 	

\safemath{\dirac}{\delta}					
\safemath{\krond}{\dirac}					

\safemath{\upto}{\uparrow}
\safemath{\downto}{\downarrow}
\safemath{\iu}{j}							
\safemath{\ev}{\lambda}						
\safemath{\hilseqspace}{l^{2}}				
\newcommand{\banachfunspace}[1]{\setL^{#1}}	
\safemath{\hilfunspace}{\banachfunspace{2}}	

\safemath{\SNR}{\textsf{SNR}} 				
\safemath{\PAR}{\textsf{PAR}} 				
\safemath{\No}{N_0}							
\safemath{\Es}{E_s}							
\safemath{\Eb}{E_b}							
\safemath{\EbNo}{\frac{\Eb}{\No}}
\safemath{\EsNo}{\frac{\Es}{\No}}

\DeclareMathOperator{\CHop}{\ensuremath{\opH}} 
\safemath{\tvir}{\rndh_{\CHop}}				
\safemath{\tvtf}{\rndl_{\CHop}}				
\safemath{\spf}{\rnds_{\CHop}}				
\safemath{\bff}{H_{\CHop}}					

\safemath{\ircf}{r_{h}}						
\safemath{\tftvcf}{r_{s}}					
\safemath{\tfcf}{r_{l}}						
\safemath{\bfcf}{r_{H}}						

\safemath{\tcorr}{c_h}						
\safemath{\scf}{c_{s}}						
\safemath{\tfcorr}{c_{l}}					
\safemath{\fcorr}{c_{H}}						

\safemath{\mi}{I}							
\safemath{\capacity}{C}						

\safemath{\normal}{\mathcal{N}}			
\safemath{\jpg}{\mathcal{CN}}			
\safemath{\mchain}{\leftrightarrow}		

\safemath{\dB}{\,\mathrm{dB}}
\safemath{\dBm}{\,\mathrm{dBm}}
\safemath{\Hz}{\,\mathrm{Hz}}
\safemath{\kHz}{\,\mathrm{kHz}}
\safemath{\MHz}{\,\mathrm{MHz}}
\safemath{\GHz}{\,\mathrm{GHz}}
\safemath{\s}{\,\mathrm{s}}
\safemath{\ms}{\,\mathrm{ms}}
\safemath{\mus}{\,\mathrm{\text{\textmu}s}}
\safemath{\ns}{\,\mathrm{ns}}
\safemath{\ps}{\,\mathrm{ps}}
\safemath{\meter}{\,\mathrm{m}}
\safemath{\mm}{\,\mathrm{mm}}
\safemath{\cm}{\,\mathrm{cm}}
\safemath{\m}{\,\mathrm{m}}
\safemath{\W}{\,\mathrm{W}}
\safemath{\mW}{\, \mathrm{mW}}
\safemath{\J}{\,\mathrm{J}}
\safemath{\K}{\,\mathrm{K}}
\safemath{\bit}{\,\mathrm{bit}}
\safemath{\nat}{\,\mathrm{nat}}

\safemath{\define}{\triangleq}			

\safemath{\equivalent}{\sim}
\safemath{\distas}{\sim}					
\safemath{\sdiff}{\Delta}				

\safemath{\reals}{\mathbb{R}}
\safemath{\positivereals}{\reals_{+}}
\safemath{\integers}{\mathbb{Z}}
\safemath{\posint}{\integers_{+}}
\safemath{\naturals}{\mathbb{N}}
\safemath{\posnaturals}{\naturals_{+}}
\safemath{\complexset}{\mathbb{C}}
\safemath{\rationals}{\mathbb{Q}}

\newcommand*{\fancyrefapplabelprefix}{app}		
\newcommand*{\fancyrefthmlabelprefix}{thm}		
\newcommand*{\fancyreflemlabelprefix}{lem}		
\newcommand*{\fancyrefcorlabelprefix}{cor}		
\newcommand*{\fancyrefdeflabelprefix}{def}		
\newcommand*{\fancyrefproplabelprefix}{prop}		
\newcommand*{\fancyrefexmpllabelprefix}{exmpl}
\newcommand*{\fancyrefalglabelprefix}{alg}		
\newcommand*{\fancyreftbllabelprefix}{tbl}		

\frefformat{vario}{\fancyrefseclabelprefix}{Section~#1}
\frefformat{vario}{\fancyrefthmlabelprefix}{Theorem~#1}
\frefformat{vario}{\fancyreftbllabelprefix}{Table~#1}
\frefformat{vario}{\fancyreflemlabelprefix}{Lemma~#1}
\frefformat{vario}{\fancyrefcorlabelprefix}{Corollary~#1}
\frefformat{vario}{\fancyrefdeflabelprefix}{Definition~#1}
\frefformat{vario}{\fancyreffiglabelprefix}{Fig.~#1}
\frefformat{vario}{\fancyrefapplabelprefix}{Appendix~#1} 
\frefformat{vario}{\fancyrefeqlabelprefix}{(#1)}
\frefformat{vario}{\fancyrefproplabelprefix}{Proposition~#1}
\frefformat{vario}{\fancyrefexmpllabelprefix}{Example~#1}
\frefformat{vario}{\fancyrefalglabelprefix}{Algorithm~#1}

 \newtheorem{prop}{Proposition}

\safemath{\dictab}{[\,\dicta\,\,\dictb\,]}

\safemath{\ysig}{\bmy}
\safemath{\ysighat}{\hat{\ysig}}
\safemath{\ysigdim}{M}
\safemath{\xsig}{\bmx}
\safemath{\xsigdim}{N}
\safemath{\nx}{n_x}
\safemath{\zsig}{\bmz}
\safemath{\zsigdim}{\ysigdim}
\safemath{\rsig}{\bmr}
\safemath{\Adict}{\bA}
\safemath{\Adicttilde}{\widetilde{\Adict}}
\safemath{\Adictdim}{\outputdim\times\xsigdim}
\safemath{\avec}{\bma}
\safemath{\avectilde}{\tilde{\avec}}
\safemath{\Bdict}{\bB}
\safemath{\Bdicttilde}{\widetilde{\Bdict}}
\safemath{\Cdict}{\bC}
\safemath{\cvec}{\bmc}
\safemath{\Ddict}{\bD}
\safemath{\Ddictdim}{\ysigdim\times\xsigdim}
\safemath{\dvec}{\bmd}
\safemath{\Ddicttilde}{\widetilde{\bD}}
\safemath{\Bonb}{\bB}
\safemath{\bvec}{\bmb}
\safemath{\Bonbdim}{\ysigdim\times\ysigdim}
\safemath{\noise}{\bmn}
\safemath{\noisedim}{\ysigim}
\safemath{\err}{\bme}
\safemath{\errdim}{\ysigdim}
\safemath{\errset}{\setE}
\safemath{\nerr}{n_e}
\safemath{\delop}{\bP_\errset}
\safemath{\delopc}{\bP_{{\errset}^c}}

\safemath{\cplxi}{\imath}
\safemath{\cplxj}{\jmath}

\safemath{\dict}{\matD}
\safemath{\inputdim}{N}		
\safemath{\outputdim}{M}		
\safemath{\sparsity}{S}	
\safemath{\inputdimA}{{N_a}}	
\safemath{\inputdimB}{{N_b}}	
\safemath{\elemA}{{n_a}}	
\safemath{\elemB}{{n_b}}	
\safemath{\resA}{\matR_a}	
\safemath{\resB}{\matR_b}	
\safemath{\subD}{\matS} 
\safemath{\subA}{\matS_a} 
\safemath{\subB}{\matS_b} 
\safemath{\dicta}{\matA} 	
\safemath{\dictb}{\matB} 	
\safemath{\hollowS}{H}
\safemath{\hollowA}{H_a}
\safemath{\hollowB}{H_b}
\safemath{\cross}{Z}
\safemath{\coh}{\mu_d}			
\safemath{\coha}{\mu_a}			
\safemath{\cohb}{\mu_b}			
\safemath{\mubs}{\nu}	
\safemath{\cohm}{\mu_m} 
\safemath{\dictset}{\setD}	
\safemath{\dictsetp}{\dictset(\coh,\coha,\cohb)}	
\safemath{\dictsetgen}{\dictset_\text{gen}}
\safemath{\dictsetgenp}{\dictsetgen(\coh)}
\safemath{\dictsetonb}{\dictset_\text{onb}}
\safemath{\dictsetonbp}{\dictsetonb(\coh)}

\safemath{\leftside}{U}
\safemath{\rightsideA}{R_a}
\safemath{\rightsideB}{R_b}

\safemath{\indexS}{\setI_S} 

\safemath{\na}{n_a}			
\safemath{\nb}{n_b}			
\safemath{\coeffa}{p_i}	
\safemath{\coeffb}{q_j}	
\safemath{\seta}{\setP}		
\safemath{\setb}{\setQ}     
\safemath{\setw}{\setW}	
\safemath{\setz}{\setZ}	
\safemath{\cola}{\veca}		
\safemath{\colb}{\vecb}		
\safemath{\cold}{\vecd}		
\safemath{\inputvec}{\vecx} 	
\safemath{\error}{\vece}	
\safemath{\noiseout}{\vecz} 	
\safemath{\inputvecel}{x}
\safemath{\inputveca}{\vecx_a}
\safemath{\inputvecb}{\vecx_b}
\safemath{\outputvec}{\vecy}	
\safemath{\lambdamin}{\lambda_{\mathrm{min}}}

\newcommand{\pos}[1]{\lefto[#1\right]^+}
\newcommand{\normtwo}[1]{\vecnorm{#1}_2}

\newcommand{\norminf}[1]{\vecnorm{#1}_\infty}
\newcommand{\norminftilde}[1]{\vecnorm{#1}_{\widetilde\infty}}

\safemath{\elltwo}{\ell_2}
\safemath{\ellone}{\ell_1}
\safemath{\ellzero}{\ell_0}
\safemath{\ellinf}{\ell_\infty}
\safemath{\ellinftilde}{\ell_{\widetilde\infty}}
\safemath{\licard}{Z(\coh,\coha,\cohb)}
\safemath{\xsol}{\hat{x}}
\safemath{\xbord}{x_b}		
\safemath{\xstat}{x_s}		
\safemath{\xstatLone}{\tilde{x}_s}
\safemath{\order}{\mathcal{O}} 
\safemath{\scales}{\Theta} 
\safemath{\ones}{\mathbf{1}} 
\safemath{\zeroes}{\mathbf{0}} 
\safemath{\thlone}{\kappa(\coh,\cohb)} 
\safemath{\constoneA}{\delta} 
\safemath{\constoneB}{\epsilon} 
\safemath{\nlarge}{L}				   
\safemath{\sumlarge}{S_\nlarge}
\safemath{\maxlarger}{P_\nlarge}	   
\safemath{\Pzero}{\textrm{P0}}	
\safemath{\Pone}{\textrm{P1}}
\safemath{\vecfir}{\vecw}			 
\safemath{\vecsec}{\vecz}
\safemath{\elvecfir}{w}              
\safemath{\elvecsec}{z}				 
\safemath{\nlargefir}{n}
\safemath{\normout}{\gamma}
\safemath{\auxfun}{h}
\safemath{\supp}{\textrm{supp}}

\safemath{\indexa}{\ell}
\safemath{\indexb}{r}
\safemath{\indexc}{i}
\safemath{\indexd}{j}

\safemath{\project}{P}

\usepackage{color}

\newcommand{\Pinf}{\ensuremath{\text{(P-INF)}}}
\newcommand{\OBR}{\ensuremath{\textsf{OBR}}}

\begin{document}

\title{PAR-Aware Large-Scale Multi-User \\ MIMO-OFDM Downlink}

\author{Christoph~Studer,~\IEEEmembership{Member,~IEEE}, 
and Erik~G.~Larsson,~\IEEEmembership{Senior Member,~IEEE} 

\thanks{Manuscript received February 1, 2012; accepted April 12, 2012.}
\thanks{Part of this paper has been presented at the 9th International Symposium on Wireless Communication Systems (ISWCS), Paris, France, Aug.~2012 \cite{SL12conf}.}
\thanks{C.~Studer is with the Dept.~of Electrical and Computer Engineering, Rice University, Houston, TX, USA (e-mail: studer@rice.edu). E.~G.~Larsson is with the Dept.\ of Electrical Engineering, Link\"oping University, Link\"oping,
Sweden (e-mail: erik.larsson@isy.liu.se).}
\thanks{The work of C.~Studer was supported by the Swiss National Science Foundation (SNSF) under Grant~PA00P2-134155. The work of E.~G.~Larsson was supported by the Swedish Foundation for Strategic Research (SSF), the Swedish Research Council (VR), and ELLIIT.  E.~G.~Larsson is a Royal Swedish Academy of Sciences (KVA) Research Fellow supported by a grant from the Knut and Alice Wallenberg Foundation.}
 
}


\maketitle


\begin{abstract}
  We investigate an orthogonal frequency-division multiplexing
  (OFDM)-based downlink transmission scheme for large-scale multi-user
  (MU) multiple-input multiple-output (MIMO) wireless systems.
  The use of OFDM causes a high peak-to-average (power) ratio (PAR), which
  necessitates expensive and power-inefficient radio-frequency (RF)
  components at the base station.
  In this paper, we present a novel downlink transmission scheme,
  which exploits the massive degrees-of-freedom available in
  large-scale MU-MIMO-OFDM systems to achieve low PAR.
  Specifically, we propose to jointly perform MU precoding, OFDM
  modulation, and PAR reduction by solving a convex optimization
  problem.
  We develop a corresponding fast iterative truncation algorithm
  (FITRA) and show numerical results to demonstrate tremendous
  PAR-reduction capabilities.
  The significantly reduced linearity requirements eventually enable
  the use of low-cost RF components for the large-scale MU-MIMO-OFDM
  downlink.
\end{abstract}




\section{Introduction}\label{sec:intro}

\IEEEPARstart{L}{arge-scale} multiple-input multiple-output (MIMO)
wireless communication is a promising means to meet the growing
demands for higher throughput and improved quality-of-service of
next-generation multi-user (MU) wireless communication
systems~\cite{RPLLETM12}.
The vision is that a large number of antennas at the base-station (BS)
would serve a large number of users concurrently and in the same 
frequency band, but with the number of BS antennas being much larger than the
number of users \cite{Marzetta10}, say a hundred antennas serving ten
users.
Large-scale MIMO systems also have the potential to reduce the
operational power consumption at the transmitter and enable the use
of low-complexity schemes for suppressing MU interference
(MUI).
All these properties render large-scale MIMO a promising technology
for next-generation wireless communication systems.

While the theoretical aspects of large-scale MU-MIMO systems have gained
significant attention in the research community, e.g.,
\cite{Marzetta06,Marzetta10,HBD11,NLM11,RPLLETM12}, much less is known
about practical transmission schemes.  As pointed out in~\cite{ML12a}, 
practical realizations of large-scale MIMO
systems will require the use of low-cost and low-power
radio-frequency~(RF) components.
To this end, reference~\cite{ML12a} proposed a novel MU
precoding scheme for frequency-flat channels, which relies 
 on per-antenna constant-envelope (CE)
transmission to enable efficient implementation using non-linear
RF components.
Moreover, the CE precoder of \cite{ML12a} forces the
peak-to-average (power) ratio (PAR) to unity,
which is not necessarily optimal as in practice there is always a
trade-off between PAR, error-rate performance, and power-amplifier efficiency.

Practical wireless channels typically exhibit frequency-selective
fading and a low-PAR precoding solution suitable for such channels
would be desirable.  Preferably, the solution should be such that the
complexity required in each (mobile) terminal is small (due to
stringent area and power constraints), whereas heavier processing
could be afforded at the BS.  Orthogonal frequency-division
multiplexing (OFDM)~\cite{NP00} is an efficient and well-established
way of dealing with frequency-selective channels.  In addition to simplifying 
the equalization at the receiver, OFDM also
facilitates per-tone power and bit allocation, scheduling in the
frequency domain, and spectrum shaping. However, OFDM is known to
suffer from a high PAR~\cite{HL05}, which necessitates the use of
linear RF components (e.g., power amplifiers) to avoid out-of-band
radiation and signal distortions. Unfortunately, linear RF components
are, in general, more costly and less power efficient than their
non-linear counterparts, which would eventually result in exorbitant
costs for large-scale BS implementations having hundreds of antennas.
Therefore, it is of paramount importance to reduce the PAR of
OFDM-based large-scale MU-MIMO systems to facilitate corresponding
low-cost and low-power BS implementations.

To combat the challenging linearity requirements of OFDM, a plethora
of PAR-reduction schemes have been proposed for
point-to-point single-antenna and MIMO wireless systems,
e.g.,~\cite{BFH96,MH97,KJ03,KJ04,FH06,IS09,TJ10}.
For MU-MIMO systems, however, a straightforward adaptation of these
schemes is non-trivial, mainly because MU systems require
the removal of MUI using a precoder~\cite{Fischer02}.
PAR-reduction schemes suitable for the MU-MISO and MU-MIMO downlink were described in~\cite{MCS08} and \cite{SF11}, respectively, and rely on  Tomlinson-Harashima precoding. Both schemes, however, require
specialized signal processing in the (mobile) terminals (e.g., modulo reduction),
which prevents their use in conventional MIMO-OFDM systems, such as IEEE
802.11n~\cite{IEEE11n} or 3GPP LTE~\cite{3GPPLTE}.

\subsection{Contributions}

In this paper, we develop a novel downlink transmission scheme for
large-scale MU-MIMO-OFDM wireless systems, which only affects the
signal processing at the BS while leaving the processing required at
each terminal untouched.
The key idea of the proposed scheme is to exploit the excess of
degrees-of-freedom (DoF) offered by equipping the BS with a large
number of antennas and to \emph{jointly} perform MU {\underline p}recoding, OFDM
{\underline m}odulation, and {\underline P}AR reduction, referred to as PMP in the remainder of the paper.
Our contributions can be summarized as follows:
\begin{itemize}
\item We formulate PMP as a convex optimization problem,
  which jointly performs MU precoding, OFDM modulation, and PAR
  reduction at the BS. 
\item We develop and analyze a novel optimization algorithm, referred
  to as fast iterative truncation algorithm~(FITRA), which is able to
  find the solution to PMP efficiently for the (typically large)
  dimensions arising in large-scale MU-MIMO-OFDM systems.
\item We present numerical simulation results to demonstrate the
  capabilities of the proposed MU-MIMO-OFDM downlink transmission
  scheme. Specifically, we analyze the trade-offs between PAR,
  error-rate performance, and out-of-band radiation, and we present a
  comparison with conventional precoding schemes.
\end{itemize}

\subsection{Notation}
\label{sec:notation}
Lowercase boldface letters stand for column vectors and uppercase
boldface letters designate matrices. For a matrix~\bA, we denote its
transpose, conjugate transpose, and largest singular value by $\bA^T$,
$\herm{\bA}$, and $\sigma_\text{max}(\bA)$, respectively;
\mbox{$\bA^{\dagger}=\bA^H\left(\bA\bA^H\right)^{\!-1}$} stands for the
pseudo-inverse of $\bA$ and the entry in the $k$th row and
$\ell$th column is $[\bA]_{k,\ell}$.
The $M\times M$ identity matrix is denoted by $\bI_M$, the $M\times N$ all-zeros matrix by $\mathbf{0}_{M\times N}$, and $\bF_M$ refers to the  $M\times M$ discrete Fourier transform~(DFT) matrix.
The $k$th entry of a vector $\bma$ is designated by $[\veca]_k$; the Euclidean (or \elltwo) norm is denoted by $\normtwo{\bma}$, $\norminf{\bma}=\max_k\abs{[\veca]_k}$ stands for the \ellinf-norm, and the $\ellinftilde$-norm~\cite{Seethaler10} is defined as
$\norminftilde{\bma}=\max\!\big\{\norminf{\Re\{\veca\}}\!,\norminf{\Im\{\veca\}}\!\big\}$
with $\Re\{\veca\}$ and $\Im\{\veca\}$ representing the real and imaginary part of~$\veca$, respectively.
Sets are designated by upper-case calligraphic letters; the cardinality and complement of the set \setT is $\abs{\setT}$ and $\setT^c$, respectively.
%
For $x\in\reals$ we define $\pos{x}=\max\{x,0\}$.

\subsection{Outline of the Paper}
The remainder of the paper is organized as
follows. \fref{sec:preliminaries} introduces the system model and
summarizes important PAR-reduction concepts.
The proposed downlink transmission scheme is detailed in \fref{sec:priorart}
and the fast iterative truncation algorithm (FITRA) is developed in
\fref{sec:FITRA}. Simulation results are presented in
\fref{sec:simulation} and we conclude in \fref{sec:conclusions}.

\section{Preliminaries}
\label{sec:preliminaries}

We start by introducing the system model that is considered in the remainder of the paper. 
We then provide a brief overview of (linear) MU precoding schemes and,
finally, we summarize the fundamental PAR issues arising in OFDM-based
communication systems.

\subsection{System Model}
\label{sec:systemmodel}

\begin{figure*}[t]
\centering
 \includegraphics[width=1.95\columnwidth]{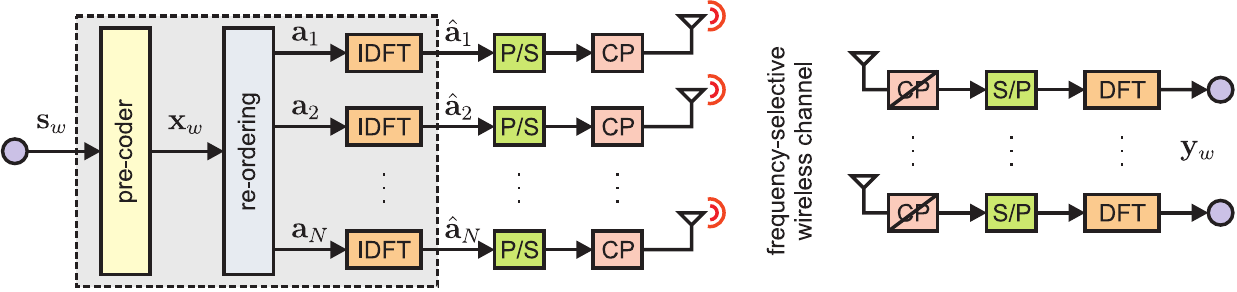}
  \caption{Large-scale MU-MIMO-OFDM downlink (left: BS with $N$ transmit antennas; right: $M$ independent single-antenna terminals). The proposed downlink transmission scheme, referred to as PMP, combines MU precoding, OFDM modulation, and PAR reduction (highlighted by the dashed box in the BS).}
  \label{fig:overview}
\end{figure*}

We consider an OFDM-based  MU-MIMO downlink scenario as depicted in \fref{fig:overview}.
The BS is assumed to have a significantly  larger number of transmit antennas~$N$ than the number $M \ll N$ of independent terminals (users); each terminal is equipped with a single antenna only.
The signal vector $\bms_w\in\setO^M$ contains information for each of
the~$M$ users, where $w=1,\ldots,W$ indexes the OFDM tones, $W$ corresponds to the total number of OFDM tones, $\setO$
represents the set of scalar complex-valued constellations,
and~\mbox{$[\bms_w]_m\in\setO$} corresponds to the symbol at tone~$w$
to be transmitted to user~$m$.\footnote{For the sake of simplicity of exposition, we employ the same constellation for all users. An extension to the general case where different constellations are used by different users is straightforward.} We normalize the symbols to satisfy
$\Exop\!\big\{\abs{[\vecs_w]_m}^2\!\big\}=1/M$.
To shape the spectrum of the transmitted signals, OFDM systems
typically specify certain unused tones (e.g., at both ends of the
spectrum~\cite{NP00}). Hence, we set $\vecs_w=\bZero_{M\times1}$ for
$w\in\setT^c$ where~$\setT$ designates the set of tones used for data
transmission.

In order to remove MUI, the signal vectors $\bms_w$, $\forall w$ are
passed through a precoder, which generates~$W$ vectors \mbox{$\bmx_w\in\complexset^{N}$}
 according to a given precoding scheme (see
\fref{sec:precoding}).
Since precoding causes the transmit power
$P=\sum_{w=1}^W\normtwo{\vecx_w}^2$ to depend on the signals
$\vecs_w$, $\forall w$ and the channel state, we normalize the
precoded vectors~$\vecx_w$, $\forall w$ prior to transmission as
\begin{align} \label{eq:renormalization}
 \hat\vecx_w = \vecx_w/\sqrt{\sum{}_{w=1}^W\normtwo{\vecx_w}^2}, \,\,  w=1,\ldots,W,
\end{align}
which ensures unit transmit power.
We emphasize that this normalization is an essential step in practice (i.e., to meet regulatory power constraints). To simplify the presentation, however, the normalization is omitted in the description of the precoders to follow (but normalization is employed in all simulation results shown in \fref{sec:simulation}). Hence, in what follows $\vecx_w$ and $\hat\vecx_w$ are treated interchangeably.

The (normalized) vectors $\vecx_w$, $\forall w$ are then re-ordered (from user orientation to transmit-antenna orientation) according to the following one-to-one mapping:
\begin{align} \label{eq:SPconversion}
\left[\,\vecx_1\,\cdots\,\vecx_{W}\,\right] = 
\left[\,\veca_1\,\cdots\,\veca_N\,\right]^T\!.
\end{align}
Here, the $W$-dimensional vector $\veca_n$ corresponds to the (frequency-domain) signal to be transmitted from the $n$th antenna.
The time-domain samples are obtained by applying the inverse DFT (IDFT) according to $\hat\veca_n = \bF^H_W \veca_n$ followed by parallel-to-serial (P/S) conversion.
Prior to modulation and transmission over the wireless channel, a cyclic prefix (CP) is added to the (time-domain) samples $\hat\veca_n$, $\forall n$ to avoid ISI~\cite{NP00}. 

To simplify the exposition, we specify the input-output relation of the wireless channel in the frequency domain only.
Concretely, we consider\footnote{We assume perfect synchronization and a CP that is longer than the maximum excess delay of the frequency-selective channel.} 
\begin{align} \label{eq:systemmodel}
\vecy_w = \matH_w\vecx_w + \vecn_w, \quad w=1,\ldots,W,
\end{align}
where $\vecy_w$ denotes the $w$th receive vector,  $\matH_w\in\complexset^{M\times{N}}$ represents the MIMO channel matrix associated with the $w$th OFDM tone, and $\vecn_w$ is an $M$-vector of i.i.d.\  complex Gaussian noise with zero-mean and variance $\No$ per entry. 
The average~receive signal-to-noise-ratio (SNR) is defined by \mbox{$\SNR=1/\No$}.
Finally, each of the $M$ user terminals performs OFDM demodulation to obtain the received (frequency-domain) signals~$[\vecy_w]_m$, $w=1,\ldots,W$ (see \fref{fig:overview}).

\subsection{MU Precoding Schemes}
\label{sec:precoding}

In order to avoid MUI, precoding must be employed at the BS.
To this end, we assume the channel matrices~$\matH_w$, $\forall w$ to be known perfectly at the transmit-side.\footnote{In large-scale MU-MIMO systems, channel-state information at the transmitter would probably be acquired through pilot-based training in the uplink and by exploiting reciprocity of the wireless channel~\cite{Marzetta10,RPLLETM12}.}
Linear precoding now amounts to transmitting $\vecx_w=\bG_w\vecs_w$, where $\bG_w\in\complexset^{N\times M}$ is a suitable precoding matrix.
One of the most prominent precoding schemes is least-squares (LS) precoding (or linear zero-forcing precoding), which corresponds to $\bG_w=\bH^\dagger_w$. Since  $\bH_w\bH^\dagger_w=\bI_M$, transmitting $\bmx_w=\bH^\dagger_w\bms_w$ perfectly removes all MUI, i.e., it transforms \fref{eq:systemmodel} into~$M$ independent single-stream systems
%
\mbox{$\vecy_w = \vecs_w + \vecn_w$}.
%
Note that LS precoding is equivalent to transmitting the solution  $\dot\vecx_w$ to the following convex optimization problem:
\begin{align*}
(\text{LS}) \quad \underset{\tilde\vecx}{\text{minimize}} \,\, \normtwo{\tilde\vecx} \quad \text{subject to}\,\, \vecs_w = \matH_w\tilde\vecx.
\end{align*}
This formulation inspired us to state the MU-MIMO-OFDM downlink transmission scheme proposed in~\fref{sec:priorart} 
 as a convex optimization problem. 

 Several other linear precoding schemes have been proposed in the
 literature, such as matched-filter (MF) precoding, minimum-mean
 square-error (MMSE) precoding~\cite{Fischer02}, or more sophisticated
 non-linear schemes, such as dirty-paper coding~\cite{EB05}.
 In the remainder of the paper, we will occasionally consider MF
 precoding, which corresponds to $\bG_w=\bH^H_w$. Since $\bH_w\bH_w^H$
 is, in general, not a diagonal matrix, MF is normally unable to
 remove the MUI. Nevertheless, MF precoding was shown in \cite{NLM11}
 to be competitive for large-scale MIMO in some operating regimes and
 in \cite{Marzetta10} to perfectly remove MUI in the large-antenna
 limit, i.e., when $N\to\infty$.

\subsection{Peak-to-Average Ratio (PAR)}
The IDFT required at the transmitter causes the OFDM signals
$\hat\veca_n$, $\forall n$ to exhibit a large dynamic
range~\cite{NP00}.
Such signals are susceptible to non-linear distortions (e.g.,
saturation or clipping) typically induced by real-world RF components.
To avoid unwanted out-of-band radiation and signal distortions
altogether, linear RF components and PAR-reduction schemes are
key to successfully deploy OFDM in practical systems.

\subsubsection{PAR Definition}
The dynamic range of the transmitted OFDM signals is typically characterized through the peak-to-average (power) ratio (PAR).
Since many real-world RF-chain implementations process and modulate
the real and imaginary part independently, we define the PAR at the
$n$th transmit antenna as\footnote{Note that alternative PAR definitions
  exist in the literature, e.g.,  using the $\ellinf$-norm in
  the nominator instead of the $\ellinftilde$-norm (and $W$ instead of
  $2W$). The relation
  $\frac{1}{2}\norminf{\hat\veca_n}^2 \leq
  \norminftilde{\hat\veca_n}^2 \leq \norminf{\hat\veca_n}^2$ shown in
  \cite[Eq.~12]{Seethaler10} ensures that reducing the PAR as defined
  in \fref{eq:PARdefinition} also reduces an $\ellinf$-norm-based PAR
  definition (and vice versa). Moreover, the theory and algorithms presented in this paper can easily be reformulated to directly reduce an $\ellinf$-norm-based PAR definition.}
\begin{align} \label{eq:PARdefinition}
\PAR_n = \frac{2W\norminftilde{\hat\veca_n}^2}{\normtwo{\hat\veca_n}^2}.
\end{align}
As a consequence of standard vector-norm relations,~\fref{eq:PARdefinition} satisfies
$1 \leq \PAR_n \leq 2W$.
%
Here, the upper bound corresponds to the worst-case PAR and is achieved for signals having only a single (real or imaginary) non-zero entry. 
The lower bound corresponds to the best case and is realized by transmit vectors whose (real and imaginary) entries have constant modulus. 
To minimize distortion due to hardware non-linearities, the
transmit signals should have a PAR that is close to one; this can
either be achieved by CE transmission~\cite{ML12a} or by using
sophisticated PAR-reduction schemes.

\subsubsection{PAR-Reduction Schemes for OFDM}
Prominent PAR-reduction schemes for single-antenna communication systems are selected mapping (SM)~\cite{BFH96}, partial transmit sequences~\cite{MH97}, active constellation extension (ACE)~\cite{KJ03}, and tone reservation (TR)~\cite{KJ04,IS09}.
PAR-reduction schemes for point-to-point MIMO systems mostly rely on SM or ACE and have been described in, e.g., \cite{FH06,TJ10}. 
For the MU-MIMO downlink, a method relying on Tomlinson-Harashima precoding and lattice reduction has been introduced recently in~\cite{SF11}; this method, however, requires dedicated signal-processing algorithms at both ends of the wireless link (e.g., modulo reduction in the receiver).
In contrast, the transmission scheme developed next aims at reducing
the PAR by \emph{only} exploiting the excess of transmit antennas
available at the BS. This approach has the key advantage of being \emph{transparent} to the receivers, i.e., it does not require any special signal-processing algorithms in the (mobile) terminals. Hence, the
proposed precoding scheme can be deployed in existing MIMO-OFDM systems for which channel-state information is
available at the transmitter, such as IEEE 802.11n~\cite{IEEE11n}.

\section{Downlink Transmission Scheme}
\label{sec:priorart}


%
The main idea of the downlink transmission scheme developed next is to
\emph{jointly} perform MU precoding, OFDM modulation, and PAR
reduction, by exploiting the DoF available in large-scale MU-MIMO
systems.
To convey the basic idea and to characterize its fundamental properties, we start by considering a simplified MIMO system. 
We then present the MU-MIMO-OFDM downlink transmission scheme in full detail and conclude by discussing possible extensions. 

\subsection{Basic Idea and Fundamental Properties}
\label{sec:basicproperties}

To convey the main idea of the proposed precoding method, let us consider an OFDM-free (narrow-band, flat-channel) MU-MIMO system with the real-valued input-output relation $\vecy=\bH\vecx+\vecn$ and an $M\times N$ channel matrix satisfying $M<N$. 
To eliminate MUI, the  transmit-vector~$\vecx$ must satisfy the precoding constraint $\vecs=\bH\vecx$, which ensures that $\vecy=\vecs+\vecn$ when transmitting the vector $\vecx$.
Since $M<N$, the equation $\vecs=\bH\vecx$ is \emph{underdetermined}; this implies that there are, in general, infinitely many solutions $\vecx$ satisfying the precoding constraint.
Our hope is now to find a suitable vector~$\dot\vecx$ having a small dynamic range (or low PAR). 

A straightforward approach that reduces the dynamic range is to transmit the solution~$\dot\vecx$ of the following optimization problem:
\begin{align*} 
(\text{P-DYN}) \,\,
\left\{\begin{array}{ll}
\underset{\alpha,\beta,\tilde\vecx}{\text{minimize}} & \alpha-\beta \\[0.05cm]
\text{subject to} & \vecs = \matH\tilde\vecx, \,\, \beta\leq\abs{[\tilde\vecx]_i}\leq\alpha,\forall i.
\end{array}\right.
\end{align*}
Unfortunately, the second constraint $ \beta\leq\abs{[\tilde\vecx]_i}\leq\alpha, \forall i$ causes this problem to be non-convex and hence, finding the solution of (P-DYN) with efficient algorithms seems to be difficult.

\subsubsection{Convex Relaxation}
To arrive at an optimization problem that reduces the dynamic range and can be solved efficiently, we relax $(\text{P-DYN})$. 
Specifically, $\beta\leq\abs{[\tilde \bmx]_i}\leq\alpha$ is replaced by $\abs{[\tilde{\bmx}]_i}\leq\alpha$, which leads to the following \emph{convex} optimization problem:
\begin{align*}
\Pinf \quad \underset{\tilde\vecx}{\text{minimize}} \,\, \norminf{\tilde\vecx} \quad \text{subject to}\,\, \vecs = \matH\tilde\vecx.
\end{align*}
Intuitively, as $\Pinf$ minimizes the magnitude of the largest entry of $\tilde\vecx$, we can expect that its solution $\dot\vecx$ exhibits low PAR.
In fact, $Ä\Pinf$ has potentially smaller PAR than a transmit vector resulting from LS precoding. 
To see this, we note that $\norminf{\dot\vecx}\leq\norminf{\bH^\dagger\vecs}$, where~$\dot\vecx$ is the minimizer of (P-INF) and $\bH^\dagger\vecs$ corresponds to the LS-precoded vector. Since~$\bH^\dagger\vecs$ is the \elltwo-norm minimizer, we have $\normtwo{\bH^\dagger\vecs}\leq\normtwo{\dot\vecx}$ and, consequently, the PAR-levels of (P-INF) and of LS precoding satisfy
\begin{align*}
\PAR_{\text{P-INF}}=\frac{N\norminf{\dot\vecx}^2}{\normtwo{\dot\vecx}^2} \leq \frac{N\norminf{\bH^\dagger\vecs}^2}{\normtwo{\bH^\dagger\vecs}^2} = \PAR_\text{LS},
\end{align*}
which implies that the PAR associated with $\Pinf$ cannot be larger than that of LS precoding. 
We confirm this observation in \fref{sec:simulation}, where the proposed downlink transmission scheme is shown to achieve substantially lower PAR than for LS precoding.

\subsubsection{Benefits of Large-Scale MIMO} \label{sec:benefitsoflargemimo}
To characterize the benefit of having a large number of transmit antennas at the BS on the PAR when using $\Pinf$, we first restate a central result from \cite{Fuchs11}.
\begin{prop}[\!\!{\cite[Prop.~1]{Fuchs11}}] \label{prop:kashin}
Let $\bH$ have full (column) rank and \mbox{$1\leq M<N$}. Generally\footnote{Note that  \cite[Prop.~1]{Fuchs11} implicitly excludes certain specific instances of the matrix~$\bH$, such as instances having collinear columns.}, the solution~$\dot\vecx$ to $\Pinf$ has $N-M+1$ entries with magnitude equal to $\norminf{\dot\vecx}$. The $M-1$ remaining entries might have smaller magnitude.
\end{prop}

With this result, we are able to derive the following upper bound on the PAR on the solution $\dot\vecx$ to  $\Pinf$:
\begin{align} \label{eq:PARbound}
\PAR_\text{P-INF} = \frac{N\norminf{\dot\vecx}^2}{\normtwo{\dot\vecx}^2} \leq \frac{N}{N-M+1}.
\end{align}
Here, the following inequality is an immediate consequence of \fref{prop:kashin}, i.e., we have
\begin{align*}
\normtwo{\dot\vecx}^2 & = \sum_{\setX} \norminf{\dot\vecx}^2 + \sum_{i\in\setX^c} \abs{[\dot\bmx]_i}^2 \\
& \geq   \sum_{\setX}  \norminf{\dot\vecx}^2 =  (N-M+1) \norminf{\dot\vecx}^2,
\end{align*}
where $\setX$ is the set of indices associated with the $N-M+1$ entries of~$\dot\vecx$ for which $\abs{[\dot\bmx]_i}=\norminf{\dot\vecx}$.
It is now key to realize that for a constant number of users~$M$ and in the large-antenna limit $N \to \infty$, the bound~\fref{eq:PARbound} implies that $\PAR_\text{P-INF} \to 1$.
Hence, for systems having a significantly larger number of transmit
antennas than users---as is the case for typical large-scale MU-MIMO
systems~\cite{Marzetta10,RPLLETM12,NLM11,HBD11}---a precoder that
implements $\Pinf$ is able to achieve a PAR that is arbitrarily close
to unity. This means that in the large-antenna limit of $N\to\infty$, $\Pinf$ yields
constant-envelope signals, while being able to perfectly eliminate the MUI.
%

\subsection{Joint Precoding, Modulation, and PAR Reduction (PMP)}
\label{sec:PARscheme}

The application of (P-INF) to each time-domain sample \emph{after}
OFDM modulation would reduce the PAR but, unfortunately, would \emph{no longer}
allow the equalization of ISI using conventional OFDM demodulation. In
fact, such a straightforward PAR-reduction approach would necessitate
the deployment of sophisticated equalization schemes in each terminal.
To  enable the use of conventional OFDM demodulation in the receiver, we next formulate the convex optimization problem, which jointly performs MU precoding, OFDM modulation, and  PAR reduction.

We start by specifying the necessary constraints.
In order to remove MUI, the following \emph{precoding constraints} must hold:
\begin{align} \label{eq:precodingconstraint}
\vecs_w=\bH_w\vecx_w, \,\,w\in\setT.
\end{align}
To ensure certain desirable spectral properties of the transmitted OFDM signals, the inactive OFDM tones (indexed by $\setT^c$) must satisfy the following \emph{shaping constraints}:
\begin{align} \label{eq:toneconstraint}
\bZero_{N\times1}=\vecx_w, \,\,w\in\setT^c.
\end{align}
PAR reduction is achieved similarly to $\Pinf$, with the main difference that we want to minimize the $\ellinftilde$-norm of the \emph{time-domain} samples $\hat\veca_n$, $\forall n$.
%
In order to simplify notation, we define the (linear) mapping between the time-domain samples~$\hat\veca_n$, $\forall n$, and the $w$th (frequency-domain) transmit vector $\vecx_w$ as 
$\vecx_w= f_w(\hat\veca_1,\ldots,\hat\veca_N)$, 
where the linear function \mbox{$f_w(\cdot)$} applies the DFT according to $\veca_n=\bF_W\hat\veca_n$, $\forall n$ and performs the re-ordering defined in~\fref{eq:SPconversion}.

With \fref{eq:precodingconstraint} and \fref{eq:toneconstraint}, we
are able to formulate the downlink transmission scheme as a convex
optimization problem:
\begin{align*}
(\text{PMP}) \,\,
\left\{\begin{array}{ll}
\underset{\tilde\veca_1,\ldots,\tilde\veca_N}{\text{minimize}} & \max\!\big\{\norminftilde{\tilde\veca_1},\ldots,\norminftilde{\tilde\veca_N}\!\big\} \\[0.2cm]
\text{subject to} & \vecs_w=\bH_wf_w(\tilde\veca_1,\ldots,\tilde\veca_N), \,\,w\in\setT \\[0.15cm]
& \bZero_{N\times1}=f_w(\tilde\veca_1,\ldots,\tilde\veca_N), \,\,w\in\setT^c.
\end{array}\right.
\end{align*}
The vectors $\hat\veca_n$, $\forall n$ which minimize (PMP) correspond to the time-domain OFDM samples to be transmitted from each antenna. Following the reasoning of \fref{sec:basicproperties}, we expect these vectors to have low PAR (see \fref{sec:simulation} for corresponding simulation results).
In what follows, ``PMP'' refers to the general method of jointly performing precoding, modulation, and PAR reduction, whereas ``(PMP)'' refers to the actual optimization problem stated above. 

\subsection{Relaxation of (PMP)}
\label{sec:relaxationpmp}

The high dimensionality of $(\text{PMP})$ for large-scale MIMO systems necessitates corresponding efficient optimization algorithms. 
To this end, we relax the constraints of (PMP) to arrive at an optimization problem that can be solved efficiently using the algorithm developed in \fref{sec:FITRA}.

To simplify the notation, we aggregate all time-domain vectors in $\overline\veca=[\,\hat\veca_1^T \,\, \cdots\,\, \hat\veca_N^T\,]^T$ and rewrite the constraints of (PMP) as a \emph{single} linear system of equations. 
Specifically, both constraints in (PMP) can be rewritten as $\overline\vecb=\overline\bC\overline\veca$, where the vector $\overline\vecb$ is a concatenation of $\vecs_w$, $w\in\setT$ and~$\abs{\setT^c}$ all-zeros vectors of dimension $N$;  the matrix $\overline\bC$ implements the right-hand-side of the constraints \fref{eq:precodingconstraint} and~\fref{eq:toneconstraint}, i.e., also includes the inverse Fourier transforms.\footnote{For the sake of simplicity of exposition, the actual structural details of the matrix $\overline\bC$ are omitted.}
We can now re-state (PMP) in more compact form  as
\begin{align*}
(\text{PMP}) \quad 
\underset{\overline\veca}{\text{minimize}}\,\,\norminftilde{\overline\veca} \quad 
\text{subject to}\,\, \overline\vecb=\overline\bC\overline\veca.
\end{align*}

In practice, it is desirable to relax the constraint $\overline\vecb=\overline\bC\overline\veca$. 
Firstly, from an implementation point-of-view, relaxing the constraints in  (PMP) enables us to develop an efficient algorithm (see \fref{sec:FITRA}).
Secondly, in the medium-to-low SNR regime, 
the effect of thermal noise  at the receiver is comparable to that of  MUI and out-of-band interference. Hence, relaxing the equation $\overline\vecb=\overline\bC\overline\veca$ to
$\normtwo{\overline\vecb-\overline\bC\overline\veca}\leq\eta$  does not significantly degrade the performance for small values of~$\eta$.
%
%
%
%
To develop an efficient algorithm for the large dimensions faced in large-scale MU-MIMO-OFDM systems (see \fref{sec:FITRA}), we state a relaxed version of (PMP) in Lagrangian form as
\begin{align*}
(\text{PMP-L}) \quad  \underset{\overline\veca}{\text{minimize}} \,\, \lambda\norminftilde{\overline\veca} + \normtwo{\overline\vecb-\overline\bC\overline\veca}^2, 
\end{align*}
where $\lambda>0$ is a regularization parameter. Note that (PMP-L) is an $\ellinftilde$-norm regularized LS problem and $\lambda$ allows one to trade fidelity to the constraints with the amount of PAR reduction (similarly to the parameter $\eta$); the associated trade-offs are investigated in  \fref{sec:tradeoff}.
Note that the algorithm developed in \fref{sec:FITRA} operates on real-valued variables.\footnote{A complex-valued formulation of the $\ellinf$-norm minimization algorithm proposed in \fref{sec:FITRA} is straightforward.} To this end,  (PMP) and (PMP-L) must be transformed into \emph{equivalent} real-valued problems. This transformation, however, is straightforward and we omit the details due to space limitations.

%


\subsection{Extensions of PMP}\label{sec:ext}

The basic ideas behind PMP can be extended to several other scenarios. Corresponding examples are outlined in the next paragraphs. 

\subsubsection{Emulating Other Linear Precoders}
By replacing the precoding constraints in \fref{eq:precodingconstraint} by
\begin{align} \label{eq:otherprecoder}
\bH_w\bP_w\vecs_w=\bH_w\vecx_w, \,\, w\in\setT,
\end{align}
where $\bP_w$ is an $N\times M$ precoding matrix of choice, one can generalize PMP to  a variety of linear precoders.
We emphasize that this generalization allows one to trade MUI removal with noise enhancement and could be used to take into account imperfect channel-state information at the transmitter, e.g., by using a minimum mean-square error precoder (see, e.g.,~\cite{Fischer02}).
\subsubsection{Peak-Power Constrained Optimization}

Instead of normalizing the power of the transmitted vectors as in \fref{eq:renormalization}, one may want to impose a predefined  upper bound  $P_{\max}$ on the transmit power already in the optimization problem. 
To this end, an additional constraint of the form $\normtwo{\overline\veca}^2\leq P_{\max}$ could be added to (PMP), which ensures that---if a feasible solution exists---the transmit power does not exceed $P_{\max}$. 
This constraint maintains the convexity of (PMP) but requires the development of a novel algorithm, as the algorithm proposed in \fref{sec:FITRA} is unable to consider such peak-power constraints in a straightforward manner.

\subsubsection{Combining PMP with Tone-Reservation (TR)}

In \cite{IS09}, the authors proposed to combine Kashin representations~\cite{LV10,Fuchs11} with TR to reduce the PAR in OFDM-based communication systems.
The underlying idea is to obtain a time-domain signal that exhibits low PAR by exploiting the DoF offered by TR.
We emphasize that PMP can easily be combined with TR, by removing certain precoding constraints \fref{eq:precodingconstraint}. Specifically,  only a subset $\setT_d\subset\setT$ is used for data transmission; the remaining tones $\setT^c_d$ are reserved for PAR reduction. This approach offers additional DoF and is, therefore, expected to further improve the PAR-reduction capabilities of PMP.

\subsubsection{Application to Point-to-Point MIMO Systems}
The proposed transmission scheme can be used for point-to-point MIMO systems for which channel-state information is available at the transmitter, e.g., IEEE 802.11n~\cite{IEEE11n}. In such systems, MUI does not need to be removed as the MIMO detector is able to separate the transmitted data streams; hence, there is potentially  more flexibility in the choice of the precoding matrices $\bP_w$, $\forall w$, as opposed to in a MU-MIMO scenario, which requires the removal of MUI.

\subsubsection{Application to Single-Carrier Systems}
The idea of PMP, i.e., to simultaneously perform precoding, modulation, and PAR reduction, can also be adapted for single-carrier large-scale MIMO systems exhibiting~ISI. 
To this end, one might want to replace the constraints in $\Pinf$ by\footnote{Note that the exact structure of the Toeplitz matrix depends on the pre- and post-ambles of the used block-transmission scheme.} 
\begin{align*} 
\left[\begin{array}{c}
\hat\vecs_1 \\
\hat\vecs_2 \\
\vdots \\
\hat\vecs_D \\
\hat\vecs_{D+1} \\
\vdots \\
\hat\vecs_Q \\
\end{array}\right] = 
\left[\begin{array}{cccc}
\widehat\bH_1 & \bZero_{M\times{}N} & \cdots & \bZero_{M\times{}N} \\
\widehat\bH_2 & \widehat\bH_1 & \cdots & \bZero_{M\times{}N}  \\
\vdots & \vdots & \ddots & \vdots \\
\widehat\bH_D & \widehat\bH_{D-1} & \cdots & \widehat\bH_1 \\
\bZero_{M\times{}N} & \widehat\bH_D & \cdots & \widehat\bH_2 \\
\vdots & \vdots & \ddots & \vdots\\
\bZero_{M\times{}N}  & \bZero_{M\times{}N} & \cdots & \widehat\bH_D
\end{array}\right] 
\left[\begin{array}{c}
\hat\vecx_1 \\
\hat\vecx_2 \\
\vdots \\
\hat\vecx_D \\
\hat\vecx_{D+1} \\
\vdots \\
\hat\vecx_Q
\end{array}\right]
\end{align*}
and minimize the $\ellinftilde$-norm of the vector $\overline\vecx=[\,\hat\vecx_1^T \,\, \cdots\,\, \hat\vecx_Q^T\,]^T$, which contains the PAR-reduced time-domain samples to be transmitted.
The channel matrices $\widehat\bH_t$ are associated to the delay (or tap) $t=1,\ldots,D$, the information symbols are denoted by $\hat\vecs_q$, $q=1,\ldots,Q$, and $Q\geq D$ refers to the number of transmitted information symbols per block. 
Alternatively to PMP, the CE precoding scheme developed in \cite{ML12a} can also be used with the constraints given above. 
A detailed investigation of both transmission schemes is, however, left for future work.

\section{Fast Iterative Truncation Algorithm}
\label{sec:FITRA}

A common approach to solve optimization problems of the form (PMP) and (PMP-L) is to use  interior-point methods~\cite{BV04}. Such methods, however, often result in prohibitively high computational complexity for the problem sizes faced in large-scale MIMO systems.
Hence, to enable practical implementation, more efficient algorithms are of paramount importance.
While a large number of computationally efficient algorithms for the $\ellone$-norm regularized LS problem have been developed in the compressive-sensing and sparse-signal recovery literature, e.g.,~\cite{candes2008a}, efficient solvers for the $\ellinf$-norm regularized LS problem (PMP-L), however, seem to be missing. 

\subsection{Summary of ISTA/FISTA}

In this section, we summarize the framework developed in~\cite{BT09} for $\ellone$-norm-based LS, which builds the basis of the algorithm derived in \fref{sec:FITRAdetails} for solving (PMP-L).

\subsubsection{ISTA}
The goal of the iterative soft-thresholding algorithm (ISTA) developed in~\cite{BT09} is to compute the solution $\hat\vecx$ to real-valued convex optimization problems of the form 
\begin{align*}
 (\text{P})\quad \underset{\vecx}{\text{minimize}} \,\, F(\vecx) = g(\vecx) + h(\vecx),
\end{align*}
where $g(\vecx)$ is a real-valued continuous convex function that is possibly non-smooth and $h(\vecx)$ is a smooth convex function, which is continuously differentiable with the Lipschitz constant $L$. 
The resulting algorithms are initialized by an arbitrary vector $\vecx_0$.
The main ingredient of ISTA is the proximal map defined as~\cite{BT09}
\begin{align} \label{eq:proximalmap}
\!\!\bmp_L(\vecy) = \argmin_{\vecx} \!\left\{g(\vecx)\! +\! \frac{L}{2}\normtwo{\vecx\!-\!\left(\vecy\!-\!\frac{1}{L}\grad{}\!h(\vecy)\right) }^2 \right\}\!,
\end{align}
which constitutes the main iteration step defined as: 
\begin{align*}
 \vecx_k = \bmp_L(\vecx_{k-1}), \quad k=1,\ldots,K.
\end{align*}
Here,~$K$ denotes the maximum number of iterations.
We emphasize that~\fref{eq:proximalmap} has a simple closed-form solution for $\ellone$-norm regularized LS,  leading to a low-complexity first-order algorithm, i.e., an algorithm requiring i) matrix-vector multiplications and ii) simple shrinkage operations only. The first property renders ISTA an attractive solution for PMP, as the involved matrices~$\overline\bC$  and its adjoint $\overline\bC^H$ exhibit a structure that enables fast matrix-vector multiplication (see \fref{sec:relaxationpmp}).

\subsubsection{Fast Version of ISTA}
As detailed in~\cite{BT09}, ISTA exhibits sub-linear convergence, i.e., \mbox{$F(\vecx_k)-F(\vecx^*) \simeq O(1/k)$}, where $\vecx^*$ designates the optimal solution to~$(\text{P})$. In order to improve the convergence rate, a fast version of ISTA, referred to as FISTA, was developed in~\cite{BT09}. 
The main idea of FISTA is to evaluate the proximal map \fref{eq:proximalmap} with a (linear) combination of the previous two points $(\vecx_{k-1},\vecx_{k-2})$ instead of $\vecx_{k-1}$ only (see \cite{BT09} for the details), which  improves the convergence rate to $F(\vecx_k)-F(\vecx^*)\simeq O(1/k^2)$ and builds the foundation of the  algorithm for solving (PMP-L) described next.

\subsection{Fast Iterative Truncation Algorithm (FITRA)}
\label{sec:FITRAdetails}

To simplify the derivation of the first-order algorithm for solving (PMP-L), we describe the algorithm for solving the Lagrangian variant of $\Pinf$ defined as follows: 
\begin{align*}
(\text{P-INF-L}) \quad \underset{\tilde\vecx}{\text{minimize}} \,\, \lambda\norminf{\tilde\vecx} + \normtwo{\vecs-\bH\tilde\vecx}^2.
\end{align*}
First, we must compute the (smallest) Lipschitz constant $L$ for the function $h(\vecx)=\normtwo{\bms-\bH\vecx}^2$ and then, evaluate the proximal map~\fref{eq:proximalmap} for the functions $g(\vecx)=\lambda\norminf{\vecx}$ and $h(\vecx)$.

\sloppy

\subsubsection{FITRA}

The (smallest) Lipschitz constant of the gradient $\grad\! h(\vecx)$ corresponds to \mbox{$L=2\sigma^2_\text{max}(\bH)$}, which can, for example, be calculated efficiently using the power method~\cite{GV96}. 
To compute the proximal map~\fref{eq:proximalmap} for (P-INF-L), we define the auxiliary vector 
\begin{align*} 
\vecw = \vecy-\frac{1}{L}\!\grad{}\!h(\vecy) = \vecy - \frac{2}{L}\bH^T\!(\bH\vecy - \vecx)
\end{align*}
which enables us to re-write the proximal map in more compact form as
\begin{align} \label{eq:proximalmaplinfy}
\bmp_L(\vecy) = \argmin_{\tilde\vecx} \left\{ \lambda\norminf{\tilde\vecx} + \frac{L}{2}\normtwo{\tilde\vecx-\vecw }^2 \right\}.
\end{align}
Unfortunately, \fref{eq:proximalmaplinfy} does---in contrast to
$\ellone$-norm regularized LS---not have a simple closed-form solution
for (P-INF-L). Nevertheless, standard algebraic manipulations enable
us to evaluate the proximal map efficiently using the following
two-step approach:
First, we compute
\begin{align} \label{eq:waterfilling}
\alpha =  \underset{\tilde\alpha}{\argmin} \, \bigg\{ \lambda\tilde\alpha + \frac{L}{2}\sum_{i=1}^N{\big(\pos{\abs{[\vecw]_i}-\tilde\alpha}\big)}^2 \bigg\},
\end{align}
for which general-purpose scalar optimization algorithms, such as the bisection method~\cite{FMM77}, can be used.\footnote{Note that in certain situations, borrowing techniques from \cite{DSSC08} may lead to faster computation of  the proximal map \fref{eq:proximalmaplinfy}.}
Then, we apply element-wise truncation (clipping) of~$\vecw$ to the interval $[-\alpha,\alpha]$ according to
$\bmp_L(\vecx) = \mathrm{trunc}_{\alpha}(\vecw)$.
The truncation operator applied to the scalar $x\in\mathbb{R}$ is defined as 
\begin{align*} 
  \mathrm{trunc}_\alpha(x) = \min\!\big\{\! \max\{x,-\alpha\}, + \alpha \big\}.
\end{align*}
The resulting first-order algorithm, including the methods proposed in \cite{BT09}  to improve the convergence rate (compared to ISTA), is detailed in \fref{alg:FITRA} and referred to as the fast iterative truncation algorithm (FITRA).

\begin{algorithm}[tp]
\caption{Fast Iterative Truncation Algorithm (FITRA)} \label{alg:FITRA}
\begin{algorithmic}[1]
\STATE {\bf initialize} $\vecx_0\gets\bZero_{N\times{}1}$, $\vecy_1 \gets \vecx_0$, $t_1 \gets1$, $L\gets2\sigma^2_\text{max}(\bH)$ 
\FOR {$k=1,\ldots,K$ }
\STATE $\vecw \gets \vecy_k - \frac{2}{L}\bH^T(\bH\vecy_k-\vecs)$ \\[0.05cm]
\STATE $\alpha \gets \underset{\tilde\alpha}{\argmin} \Big\{ \lambda\tilde\alpha + \frac{L}{2}\sum_{i=1}^N{\big(\pos{\abs{[\vecw]_i}-\tilde\alpha}\!\big)}^2 \Big\}$ \\[0.05cm]
\STATE $\vecx_k \gets \mathrm{trunc}_{\alpha}(\vecw)$ 
\STATE $t_{k+1} \gets \frac{1}{2}\big(1+\sqrt{1+4t^2_k}\big)$ 
\STATE $\vecy_{k+1} \gets \vecx_k+\frac{t_k-1}{t_{k+1}}(\vecx_k-\vecx_{k-1})$ 
\ENDFOR 
\RETURN $\!\!\vecx_{K}$
\end{algorithmic}
\end{algorithm}

\sloppy

\subsubsection{Convergence Rate}
\label{sec:convergence}

The following proposition  is an immediate consequence of the convergence results for ISTA/FISTA in \cite[Thm.~4.4]{BT09} and characterizes the convergence rate of FITRA analytically.
\begin{prop} \label{prop:FITRAconvergence}
The convergence rate of FITRA (as detailed in \fref{alg:FITRA}) satisfies 
\begin{align*}
F(\vecx_k) - F(\vecx^*) \leq \frac{2L\normtwo{\vecx_0-\vecx^*}^2}{(k+1)^2},
\end{align*}
where  $\vecx^*$ denotes the solution to (P-INF-L), $\vecx_k$ is the FITRA estimate at iteration $k$, $\vecx_0$ the initial value at iteration $k=0$, and $F(\vecx)=\lambda\norminf{\vecx} + \normtwo{\bms-\bH\vecx}^2$.
\end{prop}

We emphasize that continuation strategies, e.g., \cite{HYZ07}, potentially reduce the computational complexity of FITRA; the investigation of such methods is left for future work.


\fussy
    
\subsection{Related Work}

An algorithm to compute an approximation to (P-INF) relying an iterative truncation procedure similar to FITRA was proposed in \cite{LV10}. 
The main differences between these algorithms are as follows: The algorithm in \cite{LV10}  requires the matrix~$\bH$ to be a tight frame and relies on a constant (and pre-defined) truncation parameter, which depends on $\bH$ and cannot  be computed efficiently in practice.
In the present application, however, the matrix~$\bH$ is, in general, not a tight frame and depends on the channel realization; this requires to chose the truncation parameter in~\cite{LV10} heuristically and hence, convergence of this method is no longer guaranteed.
FITRA, in contrast, does not require the matrix~$\bH$ to be a tight frame, avoids manual tuning of the truncation parameter, and is guaranteed to converge to the solution of (P-INF-L). 
%

\section{Simulation Results}
\label{sec:simulation}

\begin{figure*}[tbp]
\centering
\subfigure[]{\includegraphics[width=0.95\columnwidth]{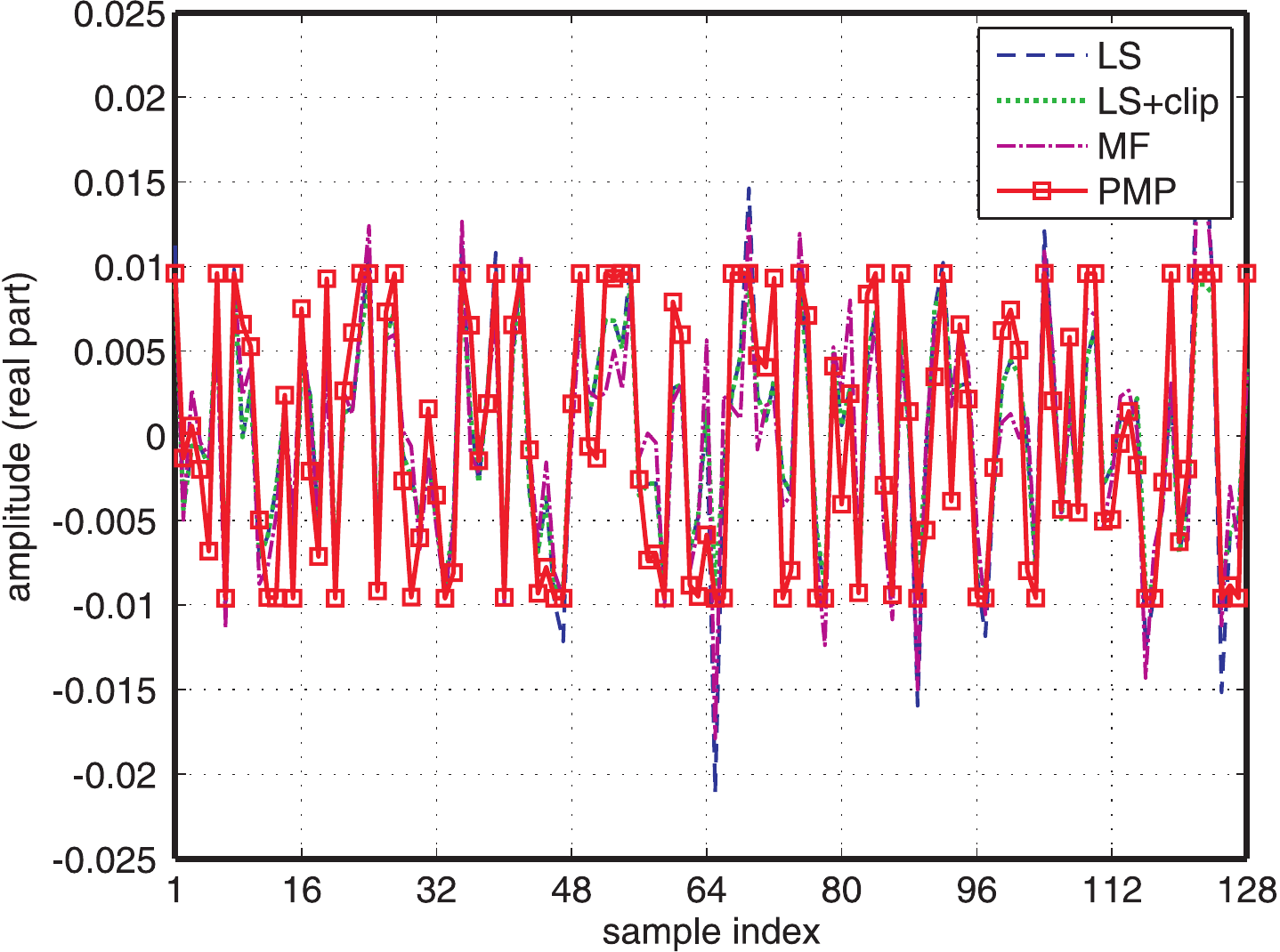}\label{fig:PARexamplea}}
\hspace{0.3cm}
\subfigure[]{\includegraphics[width=0.95\columnwidth]{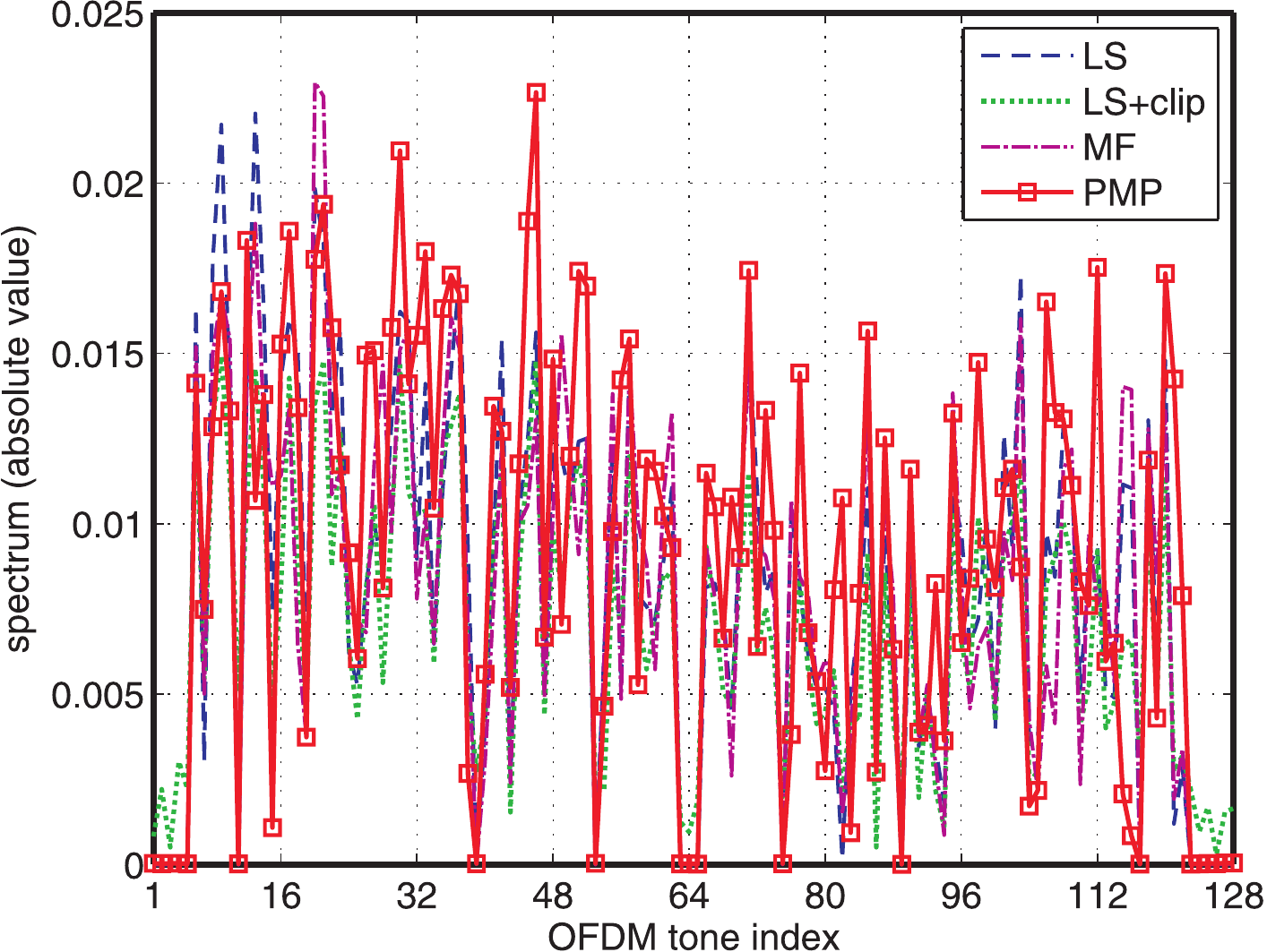}\label{fig:PARexampleb}} \\
\caption{Time/frequency representation for different precoding schemes. The target PAR for LS+clip is $4$\,dB and $\lambda=0.25$ for PMP relying on~FITRA.
(a) Time-domain signals (PAR: LS $=10.4$\,dB, LS+clip $=4.0$\,dB,  MF $=10.1$\,dB, and PMP $=1.9$\,dB).
Note that PMP generates a time-domain signal of substantially smaller PAR than LS and MF. 
(b) Frequency-domain signals (OBR: LS $=-\infty$\,dB, LS+clip $=-11.9$\,dB,  MF $=-\infty$\,dB, and PMP $=-52.9$\,dB). Note that LS, MF, and PMP preserve the spectral properties. LS+clip suffers from substantial OBR (visible at both ends of the spectrum).
}
\label{fig:PARexample1}
\end{figure*}

\begin{figure*}[tb]
\centering
\subfigure[]{\includegraphics[width=0.95\columnwidth]{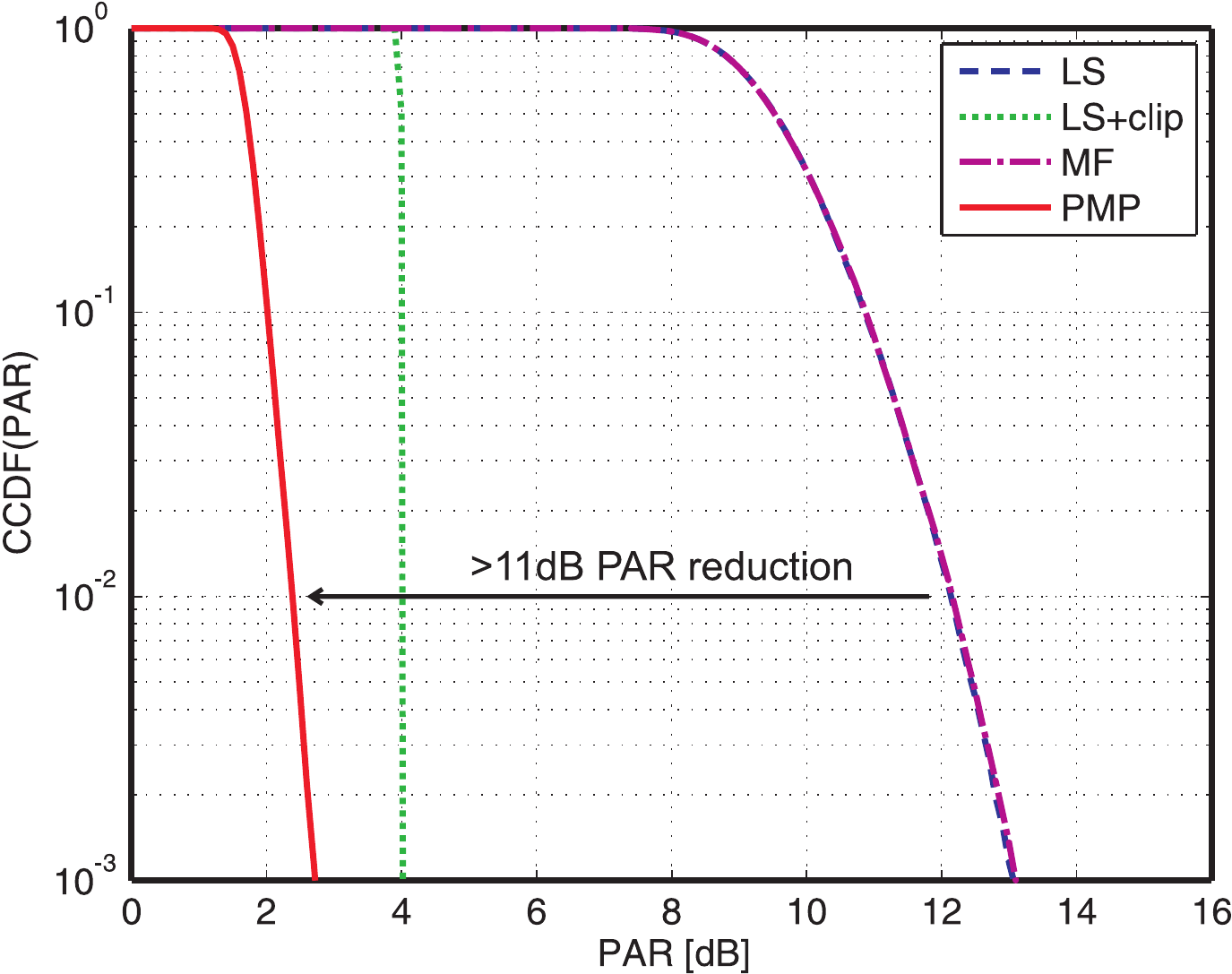}\label{fig:PARexamplec}}
\hspace{0.3cm}
\subfigure[]{\includegraphics[width=0.95\columnwidth]{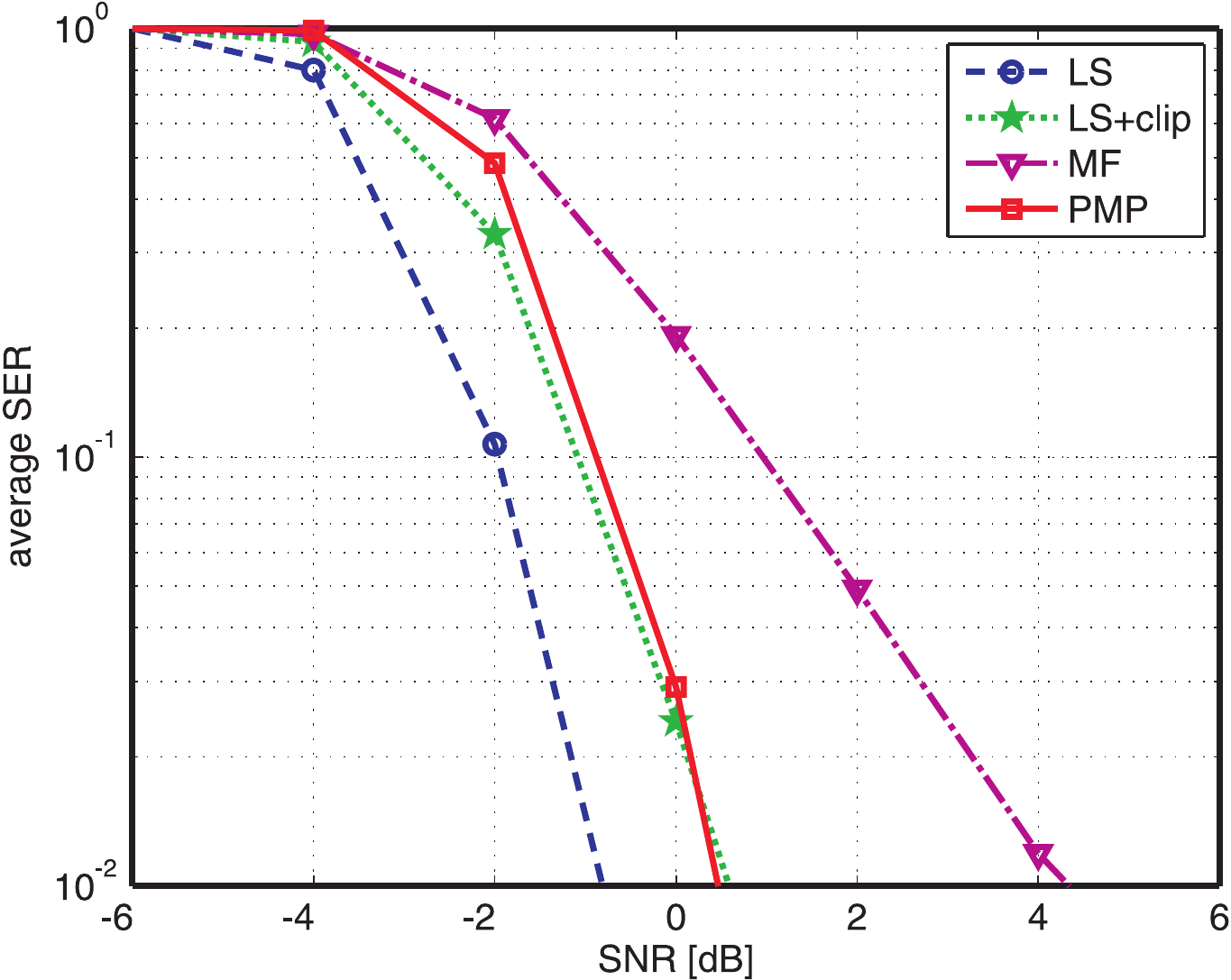}\label{fig:PARexampled}}
\caption{PAR and SER performance for various precoding schemes. The target PAR for LS+clip is 4\,dB and $\lambda=0.25$ for PMP relying on~FITRA.
(a) PAR performance (the curves of LS and MF overlap).  Note that PMP effectively reduces the PAR compared to LS and MF precoding.
(b) Symbol error-rate (SER) performance. Note that the signal normalization  causes $1$\,dB SNR-performance loss for PMP compared to LS precoding. The loss of MF is caused by residual MUI; the loss of LS+clip is caused by normalization and residual MUI.
}
\label{fig:PARexample2}
\end{figure*}

In this section, we demonstrate the efficacy of the proposed joint
precoding, modulation, and PAR reduction approach, and provide a
comparison to conventional MU precoding schemes.
 
\subsection{Simulation Parameters}

Unless explicitly stated otherwise, all simulation results are for a MU-MIMO-OFDM system having $N=100$ antennas at the BS and serving $M=10$ single-antenna terminals. 
We employ OFDM with $W=128$ tones and use a spectral map $\setT$ as specified in the 40\,MHz-mode of IEEE 802.11n~\cite{IEEE11n}.\footnote{We solely consider $\abs{\setT}=108$ data-carrying tones; the tones reserved for pilot symbols in IEEE 802.11n \cite{IEEE11n} are ignored in all simulations.}  
We consider coded transmission, i.e., for each user, we independently encode 216 information bits using a convolutional code (rate-$1/2$, generator polynomials [$133_o\,\,171_o$], and constraint length~7), apply random interleaving (across OFDM tones), and map the coded bits to a \mbox{16-QAM} constellation (using Gray labeling).

To implement (PMP-L), we use FITRA as detailed in \fref{alg:FITRA} with a maximum number of $K=2000$ iterations and a regularization parameter of $\lambda=0.25$.
In addition to LS and MF precoding, we also consider the performance of a baseline precoding and PAR-reduction method. To this end, we employ LS precoding followed by truncation (clipping) of the entries of the time-domain samples $\hat\veca_n$, $\forall n$. We use a clipping strategy where one can specify a target PAR, which is then used to compute a clipping level for which the PAR in~\fref{eq:PARdefinition} of the resulting time-domain samples is no more than the chosen target PAR.

The precoded and normalized vectors are then transmitted over a frequency-selective channel modeled as a tap-delay line with $T=4$ taps. The time-domain  channel matrices~$\widehat\bH_t$, $t=1,\ldots,T$, that constitute the impulse response of the channel, have  i.i.d.\ circularly symmetric Gaussian distributed entries with zero mean and unit variance. 
To detect the transmitted information bits, each user $m$ performs soft-output demodulation of the received symbols $[\vecy_w]_m$, $w=1,\ldots,W$ and applies a soft-input Viterbi decoder.

\subsection{Performance Measures}

To compare the PAR characteristics of different precoding schemes, we use the complementary cumulative distribution function (CCDF) defined as
\begin{align*}
\text{CCDF}(\PAR) = \Prob\{\PAR_n>\PAR\}.
\end{align*}
We furthermore define the ``PAR performance'' as the maximum PAR level $\PAR^*$ that is met for 99\% of all transmitted OFDM symbols, i.e., given by $\textsf{CCDF}(\PAR^*)=1$\%.   
The error-rate performance is measured by the  average (across users) symbol-error rate (SER); a symbol is said to be in error if at least one of the information bits per received OFDM symbol is decoded in error. The ``SNR operating point'' corresponds to the minimum SNR required to achieve 1\% SER. 
In order to characterize the amount of signal power that is transmitted outside the active tones~$\setT$, we define the out-of-band (power) ratio (OBR) as follows:
\begin{align*}
 \OBR=\frac{\abs{\setT}\sum_{w\in\setT^c}\normtwo{\vecx_w}^2}{\abs{\setT^c}\sum_{w\in\setT}\normtwo{\vecx_w}^2}.
\end{align*}
Note that for LS and MF precoding, we have $\OBR=0$, as they operate independently on each of the $W$ tones; for PMP or LS followed by clipping, we have $\OBR>0$ in general.

\subsection{Summary of PMP Properties}
\label{sec:PMPsummary}

\begin{figure*}[tb]
\centering
\subfigure[]{\includegraphics[width=0.95\columnwidth]{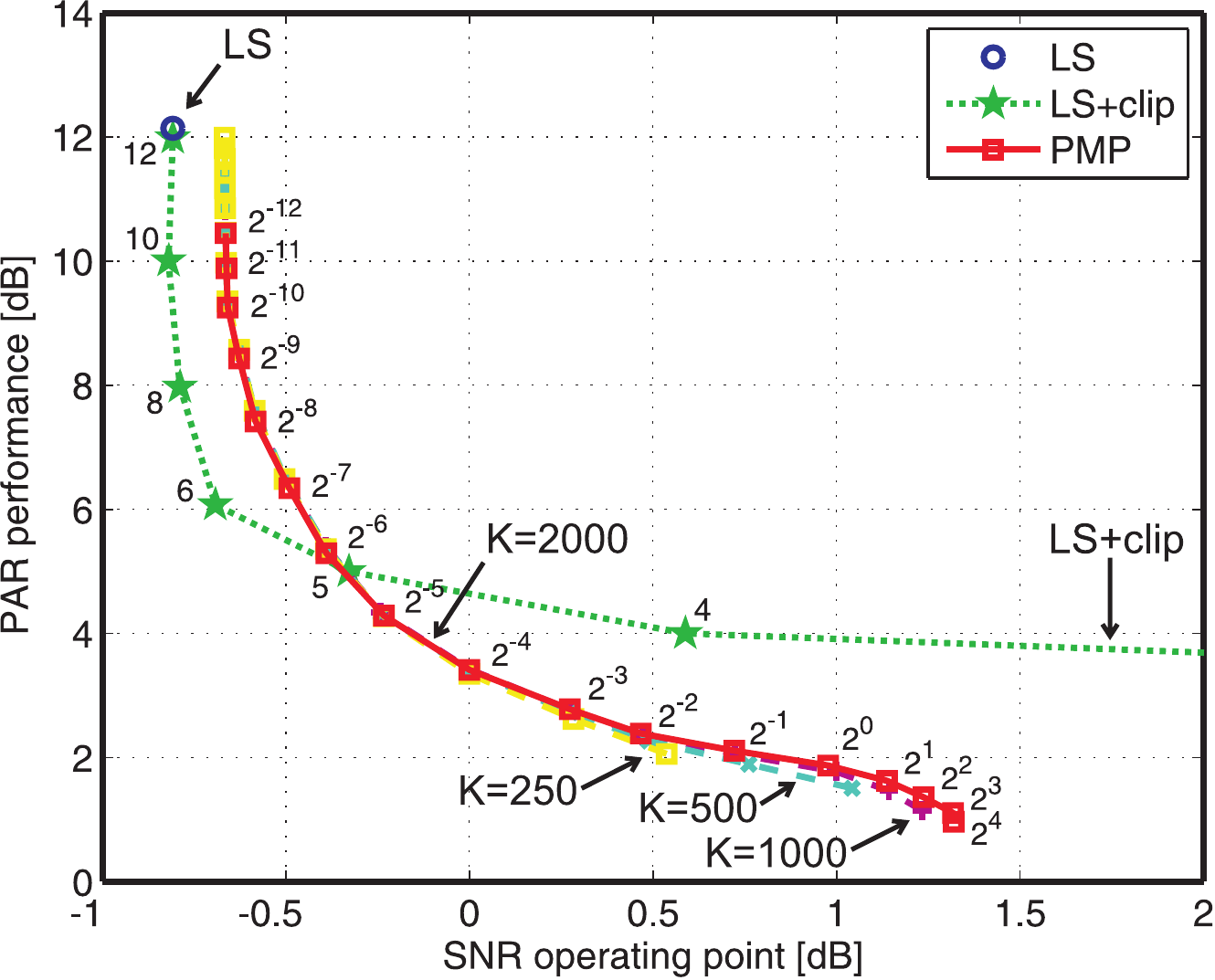}\label{fig:tradeoff_PARvsSNR}}
\hspace{0.3cm}
\subfigure[]{\includegraphics[width=0.95\columnwidth]{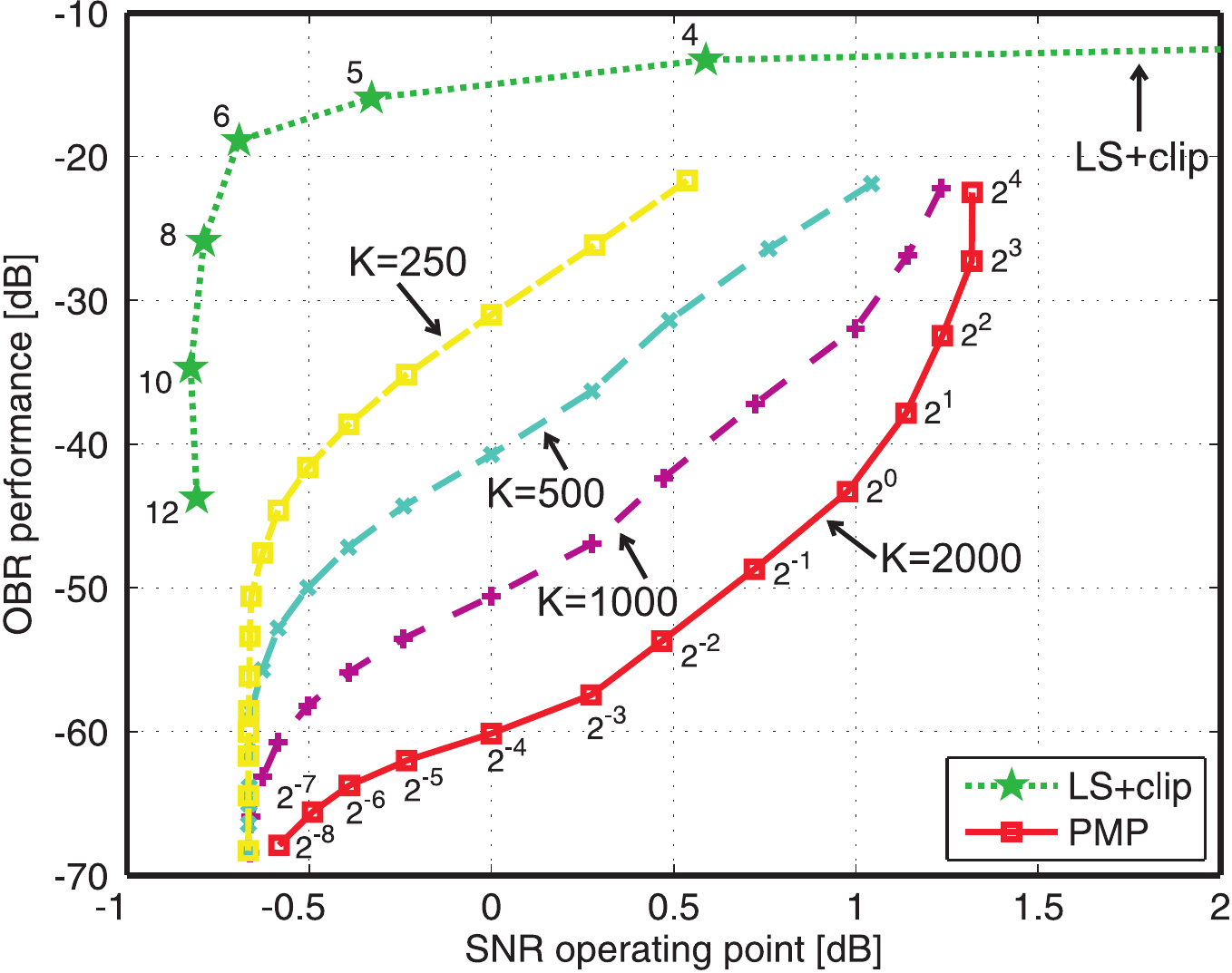}\label{fig:tradeoff_OBRvsSNR}}
\caption{SNR, PAR, and OBR performance trade-offs of PMP.  The numbers next to the trade-off curve for FITRA correspond to the regularization parameter~$\lambda$ used in (PMP-L). The LS+clip curves are parametrized by the target PAR in dB.
(a) PAR/SNR trade-off (parts of the FITRA curves overlap).
(b) OBR/SNR trade-off (all curves labeled with $K$ correspond to FITRA).}
\label{fig:tradeoff}
\end{figure*}

Figures \ref{fig:PARexample1} and \ref{fig:PARexample2} summarize the key characteristics of PMP and compare its PAR-reduction capabilities and error-rate performance to those of LS and MF precoding, as well as to LS precoding followed by clipping (denoted by ``LS+clip'' in the following).
\fref{fig:PARexamplea} shows the real part of a time-domain signal $\hat\veca_1$ for all precoding schemes (the imaginary part behaves similarly). Clearly, PMP results in time-domain signals having a significantly smaller PAR than that of LS and MF; for LS+clip the target PAR corresponds to $4$\,dB.
The frequency-domain results shown in \fref{fig:PARexampleb} confirm that LS, MF, and PMP maintain the spectral constraints. For LS+clip, however, the OBR is  $-11.9$\,dB, which is a result of ignoring the spectral constraints (see the non-zero OFDM tones at both ends of the spectrum in \fref{fig:PARexampleb}).
\fref{fig:PARexamplec} shows the PAR-performance characteristics for all considered precoding schemes. One can immediately see that PMP reduces the PAR by more than $11$\,dB compared to LS and MF precoding (at \mbox{$\textsf{CCDF}(\PAR)=1$\%}); as expected, LS+clip achieves $4$\,dB PAR deterministically.
In order to maintain a constant transmit power, the signals resulting from PMP require a stronger normalization (roughly $1$\,dB) than the signals from LS precoding; this behavior causes the SNR-performance loss compared to LS (see \fref{fig:PARexampled}).
The performance loss of MF and LS+clip is mainly caused by residual MUI.

\subsection{SNR, PAR, and OBR Trade-Offs}
\label{sec:tradeoff}

As observed in \fref{fig:PARexample2}, PMP is able to significantly reduce the PAR but results in an SNR-performance loss compared to LS precoding. 
Hence, there exists a trade-off between PAR and SER, which can be controlled by the regularization parameter $\lambda$ of (PMP-L). 
\fref{fig:tradeoff_PARvsSNR} characterizes this trade-off for $\lambda=2^v$ with $v\in\{-12,\ldots,4\}$. 
In addition to the performance of LS and MF precoding, we show the behavior of LS+clip for various target-PAR values.

 \fref{fig:tradeoff_PARvsSNR} shows that PMP is able to cover a large trade-off region that can be tuned by the regularization parameter~$\lambda$ of (PMP-L).
In particular, for a given number of FITRA iterations $K=2000$, decreasing $\lambda$ approaches the performance of LS precoding---increasing $\lambda$ reduces the PAR but results in a graceful degradation of the SNR operating point.\footnote{For $\lambda>0$, a small SNR gap remains; for $\lambda=0$, however, (PMP-L) corresponds to LS precoding and the gap vanishes.}
%
%
Hence, (PMP-L) allows one to adjust the PAR to the linearity properties of the RF components, while keeping the resulting SNR-performance loss at a minimum.
As shown in \fref{fig:tradeoff_PARvsSNR}, LS+clip achieves a similar trade-off characteristic as PMP; for less aggressive values of the target PAR, LS+clip even seems to outperform PMP.

It is important to realize that even if LS+clip outperforms PMP in terms of the PAR/SNR trade-off in the high-PAR regime, LS+clip results in substantial out-of-band interference; this important drawback is a result of ignoring the shaping constraints \fref{eq:toneconstraint}.
In particular, we can observe from \fref{fig:tradeoff_OBRvsSNR} that reducing the PAR for LS+clip quickly results in significant OBR, which renders this scheme useless in practice. By way of contrast, the OBR of PMP is significantly lower and degrades gracefully when lowering the PAR. 
Furthermore, we see that reducing the maximum number of FITRA iterations~$K$ increases the OBR. Hence, the regularization parameter $\lambda$ together with the maximum number of FITRA iterations $K$ determine the PAR, OBR, and SNR performance of PMP.
We finally note that for $K=2000$ the computational complexity of FITRA is one-to-two orders of magnitude larger than that of LS precoding. 
The underlying reason is the fact that LS precoding  solves $N$ \emph{independent} problems, whereas PMP requires the solution to a \emph{joint} optimization problem among all~$N$ transmit antennas. 

\subsection{Impact of Antenna Configuration and Channel Taps}

\begin{figure}[tb]
\centering
\includegraphics[width=0.95\columnwidth]{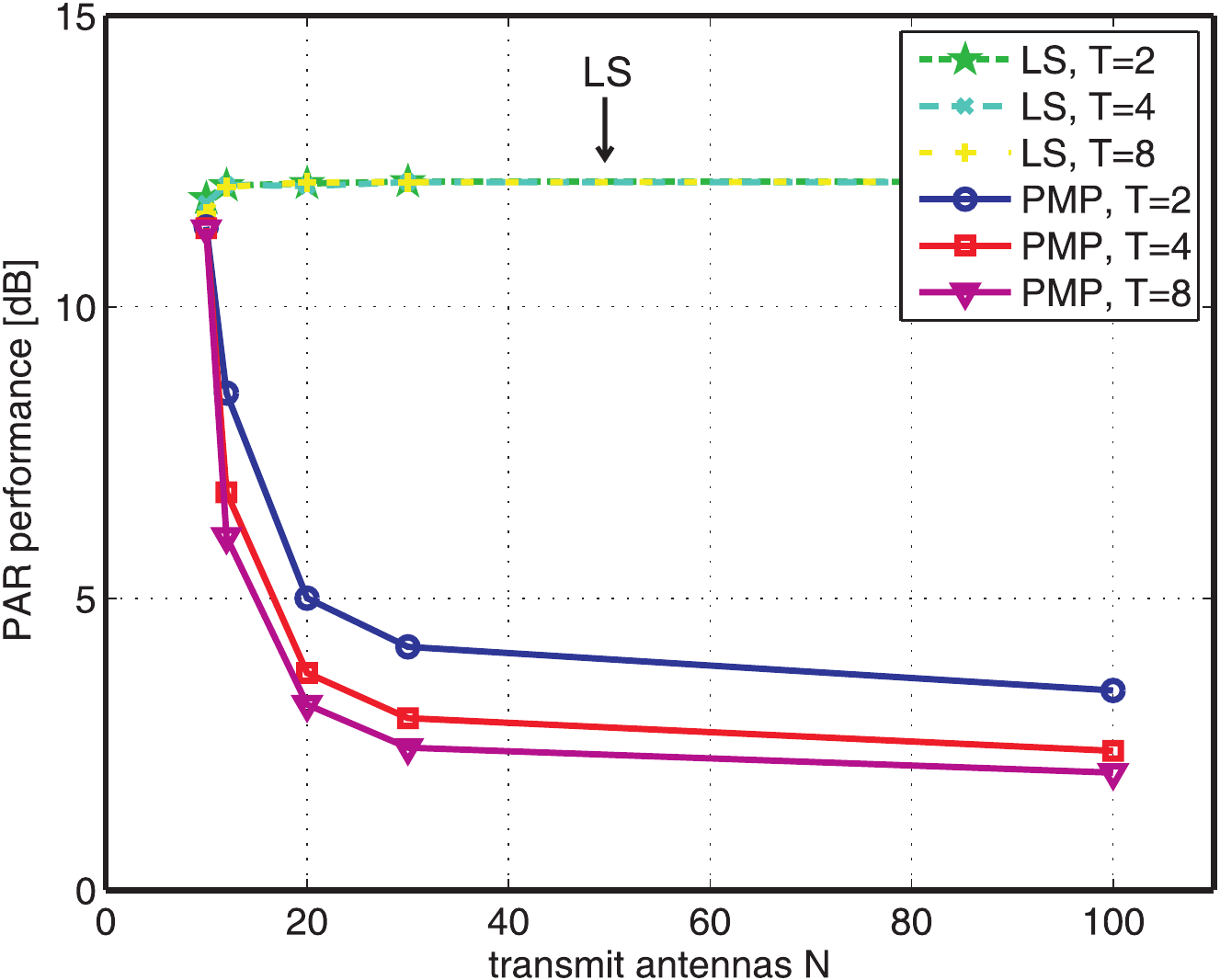}
\caption{PAR performance of PMP and LS precoding depending on the number of transmit antennas $N$ and the number of non-zero channel taps $T$; the number of users $M=10$ is held constant and $\lambda=0.25$ for PMP relying on~FITRA. (The curves for LS precoding overlap.)}
\label{fig:systempars}
\end{figure}

We finally investigate the impact of the antenna configuration to the PAR performance of PMP and LS precoding. To illustrate the impact of the channel model, we also vary the number of non-zero channel taps $T\in\{2,4,8\}$. 
\fref{fig:systempars} shows that increasing the number of transmit antennas yields improved PAR performance for PMP; this behavior was predicted analytically in~\fref{eq:PARbound} for the (narrow-band) system considered in \fref{sec:benefitsoflargemimo}.
Increasing the number of channel taps $T$ also has a beneficial impact on the PAR if using PMP. An intuitive explanation for this behavior is  that having a large number of taps increases the number of DoF, which can  then be  exploited by PMP to reduce the PAR. For LS precoding, however, the resulting PAR is virtually independent of the number of channel taps.\footnote{MF and LS+clip exhibit the same behavior; the corresponding curves are omitted in  \fref{fig:systempars}.}
In summary, PMP is suitable for MU-MIMO systems offering a large number of DoF, but also enables substantial PAR reduction for small-scale MIMO systems and channels offering only a small amount of frequency-diversity.

\section{Conclusions and outlook}
\label{sec:conclusions}

The proposed joint precoding, modulation, and PAR reduction framework, referred to as PMP,
facilitates an explicit trade-off between PAR, SNR
performance, and out-of-band interference for the large-scale
MU-MIMO-OFDM downlink.  As for the constant-envelope
precoder in \cite{ML12a}, the fundamental motivation of PMP is the
large number of DoF offered by systems where the
number of BS antennas is much larger than the number of
terminals (users).  Essentially, the downlink channel
matrix has a high-dimensional null-space, which enables us to design transmit signals
with ``hardware-friendly'' properties, such as low PAR. 
In particular, PMP yields per-antenna constant-envelope
OFDM signals in the large-antenna limit, i.e., for $N \to \infty$.
PMP is formulated
as a convex optimization problem for which a novel
efficient numerical technique, called the fast iterative truncation
algorithm (FITRA), was devised.

Numerical experiments showed that PMP is able to reduce the PAR by more than
11\,dB compared to conventional precoding methods, without creating
significant out-of-band interference; this substantially alleviates
the linearity requirements of the radio-frequency (RF) components. 
Furthermore, PMP only affects the signal processing at the BS and can
therefore be deployed in existing MIMO-OFDM wireless communication systems, such as IEEE 802.11n~\cite{IEEE11n}.

\sloppy

In addition to the extensions outlined in Section~\ref{sec:ext}, there
are many possibilities for future work. Analytical PAR-performance
guarantees of PMP are missing; the development of such
results is challenging and part of ongoing work~\cite{SYB12}.
Moreover, a detailed analysis of the impact of imperfect channel state information on the performance of PMP is left for future work.
Finally, further reducing the computational complexity of FITRA, e.g., using  continuation strategies~\cite{HYZ07}, is vital for a practical realization of PMP in hardware.
 
\section*{Acknowledgments}
 
The authors would like to thank R.~G.~Baraniuk, E.~Karipidis, A.~Maleki, S.~K.~Mohammed, G.~Pope, and A.~C.\ Sankaranarayanan for inspiring discussions. We also thank the anonymous reviewers for their valuable comments, which helped to improve the exposition of our results.

\bibliographystyle{IEEEtran} 

\bibliography{IEEEabrv,confs-jrnls,publishers,studer}

\begin{thebibliography}{10}
\providecommand{\url}[1]{#1}
\csname url@samestyle\endcsname
\providecommand{\newblock}{\relax}
\providecommand{\bibinfo}[2]{#2}
\providecommand{\BIBentrySTDinterwordspacing}{\spaceskip=0pt\relax}
\providecommand{\BIBentryALTinterwordstretchfactor}{4}
\providecommand{\BIBentryALTinterwordspacing}{\spaceskip=\fontdimen2\font plus
\BIBentryALTinterwordstretchfactor\fontdimen3\font minus
  \fontdimen4\font\relax}
\providecommand{\BIBforeignlanguage}[2]{{%
\expandafter\ifx\csname l@#1\endcsname\relax
\typeout{** WARNING: IEEEtran.bst: No hyphenation pattern has been}%
\typeout{** loaded for the language `#1'. Using the pattern for}%
\typeout{** the default language instead.}%
\else
\language=\csname l@#1\endcsname
\fi
#2}}
\providecommand{\BIBdecl}{\relax}
\BIBdecl

\bibitem{SL12conf}
C.~Studer and E.~G. Larsson, ``{PAR}-aware multi-user precoder for the
  large-scale {MIMO-OFDM} downlink,'' in \emph{Proc. of the 9th International
  Symposium on Wireless Communication Systems (ISWCS)}, Paris, France, August
  2012.

\bibitem{RPLLETM12}
F.~Rusek, D.~Persson, B.~K. Lau, E.~G. Larsson, O.~Edfors, F.~Tufvesson, and
  T.~L. Marzetta, ``Scaling up {MIMO}: opportunities and challenges with very
  large arrays,'' \emph{arXiv:1201.3210v1}, Jan. 2012.

\bibitem{Marzetta10}
T.~L. Marzetta, ``Non-cooperative cellular wireless with unlimited numbers of
  base station antennas,'' \emph{IEEE Trans. Wireless Comm.}, vol.~9, no.~11,
  pp. 3590--3600, Nov. 2010.

\bibitem{Marzetta06}
------, ``How much training is required for multi-user {MIMO}?'' in \emph{Proc.
  40th Asilomar Conf. on Signals, Systems, and Computers}, Pacific Grove, CA,
  USA, Oct. 2006, pp. 359--363.

\bibitem{HBD11}
J.~Hoydis, S.~ten Brink, and M.~Debbah, ``Massive {MIMO}: How many antennas do
  we need?'' in \emph{Proc. IEEE 49th Ann. Allerton Conf. on Comm. Control, and
  Computing (Allerton)}, Monticello, IL, USA, Sept. 2011, pp. 545--550.

\bibitem{NLM11}
H.~Q. Ngo, E.~G. Larsson, and T.~L. Marzetta, ``Energy and spectral efficiency
  of very large multiuser {MIMO} systems,'' \emph{arXiv:1112.3810v1}, Dec.
  2011.

\bibitem{ML12a}
S.~K. Mohammed and E.~G. Larsson, ``Per-antenna constant envelope precoding for
  large multi-user {MIMO} systems,'' \emph{arXiv:1111.3752v1}, Jan. 2012.

\bibitem{NP00}
R.~van Nee and R.~Prasad, \emph{{OFDM} for wireless multimedia
  communications}.\hskip 1em plus 0.5em minus 0.4em\relax Artech House Publ.,
  2000.

\bibitem{HL05}
S.~H. Han and J.~H. Lee, ``An overview of peak-to-average power ratio reduction
  techniques for multicarrier transmission,'' \emph{IEEE Wireless Comm.},
  vol.~12, no.~2, pp. 1536--1284, Apr. 2005.

\bibitem{BFH96}
R.~W. B\"auml, R.~F.~H. Fischer, and J.~B. Huber, ``Reducing the
  peak-to-average power ratio of multicarrier modulation by selected mapping,''
  \emph{IEE Elec. Letters}, vol.~32, no.~22, pp. 2056--2057, Oct. 1996.

\bibitem{MH97}
S.~H. M\"uller and J.~B. Huber, ``{OFDM} with reduced peak-to-average power
  ratio by optimum combination of partial transmit sequences,'' \emph{IEE Elec.
  Letters}, vol.~33, no.~5, pp. 368--369, Feb. 1997.

\bibitem{KJ03}
B.~S. Krongold and D.~L. Jones, ``{PAR} reduction in {OFDM} via active
  constellation extension,'' in \emph{IEEE Int. Conf. on Acoustics, Speech, and
  Sig. Proc. (ICASSP)}, vol.~4, Hong Kong, China, Apr. 2003, pp. 525--528.

\bibitem{KJ04}
------, ``An active-set approach for {OFDM} {PAR} reduction via tone
  reservation,'' \emph{IEEE Trans. Sig. Proc.}, vol.~52, no.~2, pp. 495--509,
  Feb. 2004.

\bibitem{FH06}
R.~F.~H. Fischer and M.~Hoch, ``Directed selected mapping for peak-to-average
  power ratio reduction in {MIMO} {OFDM},'' \emph{IEE Elec. Letters}, vol.~42,
  no.~2, pp. 1289--1290, Oct. 2006.

\bibitem{IS09}
J.~Illic and T.~Strohmer, ``{PAPR} reduction in {OFDM} using {Kashin's}
  representation,'' in \emph{Proc. IEEE 10th Workshop on Sig. Proc. Advances in
  Wireless Comm. (SPAWC)}, Perugia, Italy, June 2009, pp. 444--448.

\bibitem{TJ10}
T.~Tsiligkaridis and D.~L. Jones, ``{PAPR} reduction performance by active
  constellation extension for diversity {MIMO-OFDM} systems,'' \emph{J.
  Electrical and Computer Eng.}, no. 930368, 2010.

\bibitem{Fischer02}
R.~F.~H. Fischer, \emph{Precoding and Signal Shaping for Digital
  Transmission}.\hskip 1em plus 0.5em minus 0.4em\relax Wiley, New York, 2002.

\bibitem{MCS08}
S.~K. Mohammed, A.~Chockalingam, and B.~S. Rajan, ``A low-complexity precoder
  for large multiuser {MISO} systems,'' in \emph{IEEE Vehicular Tech. Conf
  (VTC)}, vol. Spring, Marina Bay, Singapore, May 2008, pp. 797--801.

\bibitem{SF11}
C.~Siegl and R.~F.~H. Fischer, ``Selected basis for {PAR} reduction in
  multi-user downlink scenarios using lattice-reduction-aided precoding,''
  \emph{EURASIP J. on Advanced Sig. Proc.}, vol.~17, pp. 1--11, July 2011.

\bibitem{IEEE11n}
\emph{{IEEE} Draft Standard; Part 11: Wireless LAN Medium Access Control (MAC)
  and Physical Layer (PHY) specifications; Amendment 4: Enhancements for Higher
  Throughput}, P802.11n/D3.0, Sep. 2007.

\bibitem{3GPPLTE}
\emph{3rd Generation Partnership Project; Technical Specification Group Radio
  Access Network; Evolved Universal Terrestrial Radio Access (E-UTRA);
  Multiplexing and channel coding (Release 9)}, 3GPP Organizational Partners TS
  36.212, Rev. 8.3.0, May 2008.

\bibitem{Seethaler10}
D.~Seethaler and H.~B\"olcskei, ``Performance and complexity analysis of
  infinity-norm sphere-decoding,'' \emph{IEEE Trans. Inf. Th.}, vol.~56, no.~3,
  pp. 1085--1105, Mar. 2010.

\bibitem{EB05}
U.~Erez and S.~ten Brink, ``A close-to-capacity dirty paper coding scheme,''
  \emph{IEEE Trans. Inf. Th.}, vol.~51, no.~10, pp. 3417--3432, Oct. 2005.

\bibitem{Fuchs11}
J.-J. Fuchs, ``Spread representations,'' in \emph{Proc. 45th Asilomar Conf. on
  Signals, Systems, and Comput.}, Pacific Grove, CA, USA, 2011.

\bibitem{LV10}
Y.~Lyubarskii and R.~Vershynin, ``Uncertainty principles and vector
  quantization,'' \emph{IEEE Trans. Inf. Th.}, vol.~56, no.~7, pp. 3491--3501,
  Jul. 2010.

\bibitem{BV04}
S.~Boyd and L.~Vandenberghe, \emph{Convex Optimization}.\hskip 1em plus 0.5em
  minus 0.4em\relax New York, NY, USA: Cambridge Univ. Press, 2004.

\bibitem{candes2008a}
E.~J. {Cand\`es} and M.~B. Wakin, ``An introdutction to compressive sampling,''
  \emph{IEEE Sig. Proc. Mag.}, vol.~25, no.~2, pp. 21--30, Mar. 2008.

\bibitem{BT09}
A.~Beck and M.~Teboulle, ``A fast iterative shrinkage-thresholding algorithm
  for linear inverse problems,'' \emph{SIAM J. Imag. Sci.}, vol.~2, no.~1, pp.
  183--202, Jan. 2009.

\bibitem{GV96}
G.~H. Golub and C.~F. {van Loan}, \emph{Matrix Computations}, 3rd~ed.\hskip 1em
  plus 0.5em minus 0.4em\relax The Johns Hopkins Univ. Press, 1996.

\bibitem{FMM77}
G.~E. Forsythe, M.~A. Malcolm, and C.~B. Moler, \emph{Computer Methods for
  Mathematical Computations}.\hskip 1em plus 0.5em minus 0.4em\relax
  Prentice-Hall, 1977.

\bibitem{DSSC08}
J.~Duchi, S.~Shalev-Shwartz, Y.~Singer, and T.~Chandra, ``Efficient projections
  onto the $\ell_1$-ball for learning in high dimensions,'' in \emph{Proc. 25th
  Int. Conf. on Machine Learning (ICML)}, Helsinki, Finland, 2008, pp.
  272--279.

\bibitem{HYZ07}
T.~Hale, W.~Yin, and Y.~Zhang, ``A fixed-point continuation method for
  $\ell_1$-regularized minimization with applications to compressed sensing,''
  Dept. Computat. Appl. Math., Rice Univ., Houston, TX, Tech. Rep. TR07-07,
  2007.

\bibitem{SYB12}
C.~Studer, W.~Yin, and R.~G. Baraniuk, ``Signal representations with minimum
  $\ell_\infty$-norm,'' in \emph{Proc. 50th Ann. Allerton Conf. on Comm.
  Control, and Computing (Allerton)}, Monticello, IL, USA, Oct. 2012.

\end{thebibliography}

\end{document}